\documentclass[a4paper,10pt,twoside]{report}
\usepackage{geometry}	
\usepackage{setspace}

\everydisplay{}

\usepackage{tocloft}
\makeatletter
\def\hlinewd#1{%
	\noalign{\ifnum0=`}\fi\hrule \@height #1 %
	\futurelet\reserved@a\@xhline}
\makeatother

\usepackage{verbatim}
 
\usepackage{hyperref}
\hypersetup{colorlinks=false}
\usepackage{array}

\usepackage[latin9,utf8]{inputenc} 
\usepackage[T1]{fontenc}
\usepackage{ae}

\usepackage{cancel}
\usepackage{float,hypcap}
\usepackage{changepage,titlesec}
\usepackage{sectsty}
\usepackage{soul}
\usepackage{bbold}
\usepackage{slashed}
\usepackage{tabularx}
\usepackage{amsmath,amsfonts,amssymb}
\usepackage{amsmath,bbm,latexsym,amssymb}
\usepackage{braket}
\usepackage{blkarray}
\usepackage{enumerate}
\usepackage{booktabs,hyperref}
\usepackage{bigints}
\usepackage{cite}
\usepackage{mflogo}
\usepackage{scrextend}
\usepackage{longtable}
\usepackage{enumitem}
\usepackage{pifont}
\usepackage{cases}
\usepackage{colortbl}
\usepackage{arydshln}

\usepackage{multirow}

\usepackage[flushleft]{threeparttable}

\usepackage{verbatim}

\usepackage{fancyvrb}
\usepackage{xcolor}

\usepackage{graphicx}
\usepackage{caption}
\usepackage{subfig}
\usepackage{feynmp-auto}

\usepackage{color}
\definecolor{orange}{rgb}{1,0.5,0}
\definecolor{uglyblue}{RGB}{95,158,160}
\definecolor{newblue}{RGB}{128,0,0}
\definecolor{mygray}{RGB}{129,129,129}
\definecolor{mywhite}{RGB}{255,250,240}
\usepackage[most]{tcolorbox}

\usepackage{authblk}

\allowdisplaybreaks

\def\be{\begin{equation}}
\def\ee{\end{equation}}
\def\ba{\begin{alignedat}}
\def\ea{\end{alignedat}}
\def\bea{\begin{eqnarray}}
\def\eea{\end{eqnarray}}
\newcommand{\bs}{\begin{subequations}}
\newcommand{\es}{\end{subequations}}

\def\vs{\vspace}

\def\no{\nonumber\\}
\def\fn{\footnote}

\newcommand{\newc}{\newcommand}
\newc{\ol}{\overline}
\newc{\wt}{\widetilde}

\newc{\m}{\mathcal}

\newcommand{\beq}{\begin{eqnarray}}
\newcommand{\eeq}{\end{eqnarray}}
\newcommand{\bpmatrix}{\begin{pmatrix}}
\newcommand{\epmatrix}{\end{pmatrix}}

\renewcommand{\ol}{\text{1l}}

\renewcommand{\eqref}[1]{eq.~(\ref{#1})}

\newcommand{\bc}{\begin{center}}
\newcommand{\ec}{\end{center}}

\def\m#1{m_{#1}}

\def\ltap{\;\centeron{\raise.35ex\hbox{$<$}}{\lower.65ex\hbox{$\sim$}}\;}
\def\gtap{\;\centeron{\raise.35ex\hbox{$>$}}{\lower.65ex\hbox{$\sim$}}\;}

\usepackage{stackengine}
\stackMath

\usepackage{url}
\usepackage{epsf, array, color}

\begin{document}

\pagestyle{plain}

\makeatletter         
\renewcommand\maketitle{
\begin{center}
{\large \bfseries \@title }

\vspace{1mm}

{\@author}

\vspace{1.2mm}

(\@date)
\end{center}} 
\makeatother

\title{Role of dimension-eight operators in an EFT for the 2HDM}
%
\renewcommand*{\thefootnote}{\fnsymbol{footnote}}
\author[1]{Sally Dawson\fn{dawson@bnl.gov}}
\author[1]{Duarte Fontes\fn{dfontes@bnl.gov}}
\author[2]{Samuel Homiller\fn{shomiller@g.harvard.edu}}
\author[1]{Matthew Sullivan\fn{msullivan1@bnl.gov}}
\affil[1]{\textit{Department of Physics, Brookhaven National Laboratory, Upton, New York 11973 U.S.A.}}
\affil[2]{\textit{Physics Department, Harvard University, Cambridge, MA, 02138, U.S.A.}}
\date{\today}

\maketitle

\renewcommand*{\thefootnote}{\arabic{footnote}}
\setcounter{footnote}{0}

\vs{0.5mm}
\begin{addmargin}[12mm]{12mm}
\small
The Standard Model effective field theory (SMEFT) is the tool of choice for studying deviations of Higgs couplings from the Standard Model predictions.  The SMEFT is an expansion in an infinite tower of higher dimension operators, which is typically truncated at dimension-6.  We consider the effective theory including dimension-8 operators and examine the matching to the 2 Higgs Doublet Model (2HDM) that is assumed to be valid at some high scale.   Both the limits from the direct production of single Higgs bosons and the indirect limits on the Higgs tri-linear coupling are considered in the context of the SMEFT matched to the 2HDM, and the importance of the assumptions about the expansions in powers of the high scale are examined numerically. 
\end{addmargin}

\normalsize

\section{Introduction}
\label{sec:intro}

One of the major goals of the LHC program is to extract information about high scale --- or ultraviolet (UV) --- physics from precision measurements of Higgs couplings. In the scenario where the only observed particles are those of the Standard Model (SM), the measurements can be analyzed in an effective field theory (EFT) framework, where the Lagrangian $\mathcal{L}$ contains an infinite tower of higher dimension operators $O_i^n$ (for a review, see Ref. \cite{Brivio:2017vri}),
\begin{equation}
\label{eq:basics}
\mathcal{L} = \mathcal{L}_{\mathrm{SM}} + \Sigma_{n,i}{C_i^n O_i^n\over 
\Lambda^{n-4}
} \, .
\end{equation}
(Here, $\mathcal{L}_{\mathrm{SM}}$ is the SM Lagrangian, $\Lambda$ is the UV scale and $C_i^n$ are coefficient functions.) 
Considering the Higgs as an SU(2) doublet defines this as the Standard Model effective field theory (SMEFT).
In this way, all new physics effects are contained in the coefficient functions $C_i^n$, also known as Wilson coefficients (WCs). These can be constrained by  the LHC Higgs and top quark data, $W^+W^-$ data at both LEP-II and the LHC,  and precision electroweak measurements, through  analyses that use increasingly sophisticated global fits\cite{Almeida:2021asy,Ethier:2021bye,deBlas:2021wap,Ellis:2020unq,Alasfar:2020mne,DeBlas:2019qco,Biekoetter:2018ypq,DiVita:2017eyz}.

The ultimate goal of this  framework is a double one: first, to observe a pattern of non-zero coefficient functions; then, to associate that pattern with a specific new physics model. In the absence of any definitive observation, one can pave the way by focusing on the second task, i.e. by studying the quality with which one can associate the EFT description to that of different models. Several studies considered the matching between the two descriptions at tree-level and truncated Eq. \ref{eq:basics} with dimension-6 operators; for example, simple cases where the UV physics arises from the exchange of a single weakly coupled particle have been categorized in the pioneering work of Ref. \cite{deBlas:2017xtg}, and the effects in some well known models such as the 2 Higgs Doublet Model (2HDM) or the Higgs singlet model were studied in Refs.  \cite{
Perez:1995dc,
Englert:2014uua,
Brehmer:2015rna,
Gorbahn:2015gxa,
Belusca-Maito:2016dqe,
Dawson:2017vgm,
Dawson:2020oco}.

However, there are many avenues that need to be explored to put this program on a solid footing. One of them investigates the effects of radiative corrections to the EFT predictions\cite{
Cullen:2020zof,
Cullen:2019nnr,
Gauld:2016kuu,
Hartmann:2015aia,
Hartmann:2015oia,
Dawson:2018liq,
Dedes:2018seb,
Dawson:2018pyl,
Dedes:2019bew,
Dawson:2019clf,
Hartmann:2016pil,
Boughezal:2019xpp,
Dawson:2018dxp,
Dawson:2021xea}.
  Furthermore, the one-loop matching between the UV theory and the EFT at the high scale  can potentially have significant numerical effects\cite{Henning:2016lyp,DasBakshi:2018vni,Cohen:2022tir,Carmona:2021xtq,Criado:2017khh}  and was perfomed in the Higgs singlet model \cite{Jiang:2018pbd,Haisch:2020ahr,Dawson:2021jcl,Cohen:2020xca,Anisha:2021hgc}, where the numerical impact of the one-loop matching ends up being small.   
In another relevant avenue, one can explore the effects of higher dimension operators. The global fits described above typically truncate the expansion with the dimension-6 operators. The numerical relevance of dimension-8 operators is not known in general, having been clarified in only a few specific cases. One of them concerns the $WWh$ vertex, where the contribution from the dimension-8 operator to $Wh$ production at the LHC is ${\cal{O}}(15\%)$ \cite{Hays:2018zze}. Yet the size of the dimension-8 effects turns out to be extremely model dependent: whereas an SU(2) vector triplet model yields significant dimension-8 effects in the analysis of precision electroweak data\cite{Corbett:2021eux}, a vector-like top partner induces negligible dimension-8 effections in $t{\overline t}h$ production\cite{Dawson:2021xei}. 

In this work, we examine the importance of dimension-8 operators in the EFT for the 2HDM at the leading order in the loop expansion. A strong motivation for this study is the fact that at dimension-6 (and for the Type-I version of the full model), the SMEFT poorly reproduces the predictions of the 2HDM \cite{Belusca-Maito:2016dqe}, since the Higgs coupling to gauge bosons first arises at dimension-8. 
We organize the paper as follows: in Section \ref{sec:2hdm}, we describe our conventions for the 2HDM in order to set the notation; we then present the matching to the SMEFT to dimension-8 in Section \ref{sec:match}, and some numerical results for fits to Higgs coupling measurements in the full 2HDM and the SMEFT in Section \ref{sec:res}; finally, we present our conclusions in Section \ref{sec:conc}.

\section{2HDM recap}
\label{sec:2hdm}

In this section, we review the 2HDM \cite{Lee:1973iz,Gunion:1989we,Branco:2011iw} (see also Refs. \cite{Belusca-Maito:2016dqe,Belusca-Maito:2017iob}). Besides the SM scalar doublet $\Phi_1$, the model contains an extra doublet $\Phi_2$; each doublet has in general a non-zero vacuum expectation value (vev), which we identify as $v_1/\sqrt{2}$ and $v_2/\sqrt{2}$, respectively. We assume a softly broken $Z_2$ symmetry, according to which $\Phi_1 \to \Phi_1, \Phi_2 \to -\Phi_2$. This symmetry is extended to the fermion fields; such an extension can be made in four different ways, each one corresponding to a different type of 2HDM: Type-I, Type-II, Type-L (or Lepton-Specific) and Type-F (or Flipped); for details, see Ref. \cite{Branco:2011iw}. 

It is convenient to work in the so-called Higgs basis\cite{Donoghue:1978cj,Georgi:1978ri,Botella:1994cs,Branco:1999fs}, where the doublets are identified as $H_{1}$ and $H_{2}$, and which is obtained by rotating the doublets in the original basis according to:
\be
\label{eq:basis-rot}
\left(\begin{array}{c}
H_{1} \\
H_{2}
\end{array}\right)=\left(\begin{array}{cc}
c_{\beta} & s_{\beta} \\
-s_{\beta} & c_{\beta}
\end{array}\right)\left(\begin{array}{c}
\Phi_{1} \\
\Phi_{2}
\end{array}\right),
\ee
where we introduce the short-hand notation $c_x=\cos(x), ~s_x=\sin(x)$, which we use for any angle $x$.
The angle $\beta$ in Eq. \ref{eq:basis-rot} obeys the relation $\tan \beta = v_2/v_1$, which implies that, in the Higgs basis, the vev is entirely contained in the first doublet; that is, the vev of $H_2$ is zero, whereas the vev $H_1$ is $v/\sqrt{2}$, with $v =\sqrt{v_1^2 + v_2^2} = 246$ GeV.
For our purposes, we can focus on three sectors of the full Lagrangian:
\be
\label{eq:2HDM-Lag}
\mathcal{L}_{\mathrm{2HDM}} \ni \mathcal{L}_{\mathrm{kin}}  + \mathcal{L}_Y -V, 
\ee
where the terms in the right-hand side respectively correspond to the scalar kinetic piece, the Yukawa piece and the potential. In the Higgs basis, these sectors can be written as:
\bs
\bea
\mathcal{L}_{\mathrm{kin}} &=& \left(D_{\mu} H_1\right)^{\dagger} \left(D^{\mu} H_1\right) + \left(D_{\mu} H_2\right)^{\dagger} \left(D^{\mu} H_2\right),\\[3mm]
\label{eq:potential}
V &=& Y_1 H_{1}^{\dagger} H_{1}
+ Y_2 H_{2}^{\dagger} H_{2}+\left(Y_3 H_{1}^{\dagger} H_{2}+\textrm{h.c.}\right) \no
&&+ \frac{Z_{1}}{2}\left(H_{1}^{\dagger} H_{1}\right)^{2}+\frac{Z_{2}}{2}\left(H_{2}^{\dagger} H_{2}\right)^{2}+Z_{3}\left(H_{1}^{\dagger} H_{1}\right)\left(H_{2}^{\dagger} H_{2}\right)+Z_{4}\left(H_{1}^{\dagger} H_{2}\right)\left(H_{2}^{\dagger} H_{1}\right) \no
&& + \left\{\frac{Z_{5}}{2}\left(H_{1}^{\dagger} H_{2}\right)^{2}+Z_{6}\left(H_{1}^{\dagger} H_{1}\right)\left(H_{1}^{\dagger} H_{2}\right)+Z_{7}\left(H_{2}^{\dagger} H_{2}\right)\left(H_{1}^{\dagger} H_{2}\right)+ \textrm{h.c.}\right\}, \\[3mm]
\mathcal{L}_Y &=&
- \lambda_u^{(1)} \bar{u}_R \tilde{H}_1^{\dagger} q_L 
- \lambda_u^{(2)} \bar{u}_R \tilde{H}_2^{\dagger} q_L
- \lambda_d^{(1)} \bar{d}_R H_1^{\dagger} q_L 
- \lambda_d^{(2)} \bar{d}_R H_2^{\dagger} q_L 
- \lambda_l^{(1)} \bar{e}_R H_1^{\dagger} l_L 
- \lambda_l^{(2)} \bar{e}_R H_2^{\dagger} l_L 
+ \textrm{h.c.} \no
&=& 
- \lambda_u^{(1)*} H_1^{\dagger} \widehat{q}_L u_R
- \lambda_u^{(2)*} H_2^{\dagger} \widehat{q}_L u_R
- \lambda_d^{(1)} \bar{d}_R H_1^{\dagger} q_L 
- \lambda_d^{(2)} \bar{d}_R H_2^{\dagger} q_L 
- \lambda_l^{(1)} \bar{e}_R H_1^{\dagger} l_L 
- \lambda_l^{(2)} \bar{e}_R H_2^{\dagger} l_L \no
&& \qquad + \textrm{h.c.},
\eea
\es
where $\tilde{H}_i = i \sigma_2 H_i^*$ and  we introduce $\widehat{q}_L \equiv -i \sigma_2 (\bar{q}_L)^{\mathrm{T}}$.%
\fn{Making the SU(2) components explicit, 
$
\widehat{q}_L
=
\begin{pmatrix}
-\bar{d}_L \\
 \bar{u}_L
\end{pmatrix}
$.
This implies the relation 
$\bar{u}_R {\widehat{q}_L}^{\dagger} H_j = \bar{u}_R {\tilde{H}_j}^{\dagger} q_L$,
as well as its Hermitian conjugate
$H_j^{\dagger} \widehat{q}_L u_R = \bar{q}_L \tilde{H}_j u_R$.}
We neglect generation indices on the right-handed SU(2) singlets $u_R,~d_R$ and $e_R$ and on the left-handed SU(2) doublets, $q_L$ and $l_L$.
The parameters $Y_3, Z_{5}, Z_{6}, Z_{7}$ and the Yukawa parameters $\lambda_f^{(i)}$ are in general complex --- for $i=1,2$, and where $f$ can represent any of the three types of charged fermions: up-type quarks ($u$), down-type quarks ($d$) and charged leptons ($l$). The remaining parameters are required to be real by hermiticity.
Note, however, that the phases of $\lambda_f^{(i)}$ can always be absorbed by the fermion fields; hence, we take $\lambda_f^{(i)}$ to be real without loss of generality.
We thus write:
\be
\label{eq:2HDM-Yukawas}
\lambda_f^{(1)} = \dfrac{\sqrt{2}}{v} m_f,
\qquad
\lambda_f^{(2)} = \dfrac{\eta_f}{\tan\beta} \lambda_f^{(1)},
\ee
where $\eta_f$ depends on the particular type of 2HDM, given in Table \ref{tab:types}, and $m_f$ is to be understood as a 3x3 diagonal matrix in the flavour space containing the masses of the three generations of fermion type $f$. %
Finally, the minimization equations read:
\be
\label{eq:theYs}
Y_1 = - \dfrac{Z_1}{2} v^2,
\qquad
Y_3 = - \dfrac{Z_6}{2} v^2 \, .
\ee
\begin{table}[h!]
\centering
\begin{tabular}
{
@{\hspace{-0.8mm}}
>{\centering}p{1cm}
>{\centering}p{1.8cm}
>{\centering}p{1.8cm}
>{\centering}p{1.8cm}
>{\centering\arraybackslash}p{1.8cm}
@{\hspace{3mm}}
}
\hlinewd{1.1pt}
& Type-I &  Type-II & Type-L & Type-F \\
\hline
$\eta_{u} $ & 1 & 1 & 1 & 1 \\
$\eta_{d}$ & 1 & $-\tan ^{2} \beta$ & 1 & $-\tan ^{2} \beta$ \\
$\eta_{l}$ & 1 & $-\tan ^{2} \beta$ & $-\tan ^{2} \beta$ & 1 \\
\hlinewd{1.1pt}
\end{tabular}
\caption{Values of the parameter $\eta_f$ for the different types of 2HDM and for the different types of charged fermions.}
\label{tab:types}
\end{table}
\normalsize

In this work, we are interested in the scenario where the parameters $Y_3, Z_{5}, Z_{6}, Z_{7}$ all take real values, in which case the scalar sector preserves CP symmetry at the leading order.%
\fn{It should be clear, however, that this limit should be seen as a particular solution of the model where the parameters are in general complex, and not as a model by itself \cite{Fontes:2021znm}.} 
Accordingly, we can parameterize $H_1$ and $H_2$ as:
\begin{align}
H_1 = 
\begin{pmatrix}
G^+ \\
\frac{1}{\sqrt{2}}(v + h_1^{\mathrm{H}} + i G_0)
\end{pmatrix},
\hspace{3mm}
H_2 = 
\begin{pmatrix}
H^+ \\
\frac{1}{\sqrt{2}}(h_2^{\mathrm{H}} + i A)
\end{pmatrix},
\label{eq:parametrizacao-Hs-real}
\end{align}
with $G^+, H^+$ complex fields, and $h_1^{\mathrm{H}}, h_2^{\mathrm{H}}, A, G_0$ real fields. All of these states are already mass eigenstates, except $h_1^{\mathrm{H}}, h_2^{\mathrm{H}}$; the states $G^+$ and $G_0$ are the would-be Goldstone bosons, and $H^+$ and $A$ correspond to the charged scalar and the pseudo-scalar bosons, respectively. The scalar mass states $h_{125}$ and $H_0$ are obtained from $h_1^{\mathrm{H}}$ and $h_2^{\mathrm{H}}$ by introducing the mixing angle $\alpha$, such that:
\be
\label{eq:the-doublets}
H_{1} = \left(\begin{array}{c}
G^{+} \\
\frac{1}{\sqrt{2}}\left(v + s_{\beta \! - \! \alpha} \, h_{125} + c_{\beta \! - \! \alpha} \, H_{0} + i G_{0}\right)
\end{array}\right),
\quad
H_{2} = \left(\begin{array}{c}
H^{+} \\
\frac{1}{\sqrt{2}}\left(c_{\beta \! - \! \alpha} \,  h_{125} - s_{\beta \! - \! \alpha} \, H_{0} + i A\right)
\end{array}\right),
\ee
where the field $h_{125}$ is identified with the 125 GeV scalar observed at the LHC.

Finally, in the CP conserving solution, the $Z_i$ parameters can be written in terms of the masses of the physical scalars, the mixing parameter $\beta \! - \! \alpha$ and the parameters $Y_2$ and $v^2$ according to:
\bs
\label{eq:theZs-real}
\bea
Z_1 &=& \dfrac{ \Big[ m_{h_{125}}^2 + m_{H_0}^2 \, \cot(\beta \! - \! \alpha)^2 \Big]  \, \sin(\beta \! - \! \alpha)^2}{v^2},
\label{eq:zmass1} \\
Z_3 &=& \dfrac{2 \,  (m_{H^{\pm}}^2 - Y_2) }{v^2}, \\
Z_4 &=& \dfrac{2 \, m_A^2 + m_{h_{125}}^2 + m_{H_0}^2 - 4 \, m_{H^{\pm}}^2 +  (m_{h_{125}}^2 - m_{H_0}^2) \, \cos\Big[2 \,  (\beta \! - \! \alpha)\Big] }{2 \, v^2}, \\
Z_5 &=& \dfrac{-2 \, m_A^2 + m_{h_{125}}^2 + m_{H_0}^2 + (m_{h_{125}}^2 - m_{H_0}^2)  \, \cos\Big[2 \,  (\beta \! - \! \alpha)\Big] }{2 \, v^2}, \\
Z_6 &=& \dfrac{ (m_{h_{125}}^2 -m_{H_0}^2) \, \sin\Big[2 \,  (\beta \! - \! \alpha)\Big] }{2 \, v^2},
\label{eq:zmass}
\eea
\es
where $m_{h_{125}}, m_{H_0}, m_A$ and $m_{H^{\pm}}$ correspond to the masses of $h_{125}, H_0, A$ and $H^{\pm}$, respectively. In our analysis, we shall take the following parameters as independent:
\bs
\bea
\label{eq:indep-real}
&
\beta \! - \! \alpha,
\,
m_{h_{125}},
\,
Y_2,
\,
m_{H_0},
\,
m_A,
\,
m_{H^{\pm}},
\ \ 
&\text{(scalar sector),} \\
&
\beta,
\,
m_f
\ \ 
&\text{(Yukawa sector)}.
\eea
\es

\section{EFT}
\label{sec:match}

\subsection{The effective Lagrangian}
\label{sec:IOH2}

We want to obtain the effective Lagrangian that results from integrating out the doublet $H_2$, which is assumed to be heavy.%
\fn{A detailed explanation of this procedure up to dimension-8 operators can also be found in Ref. \cite{Belusca-Maito:2016cay}.}
To clarify how this is done, let us suppose a generic theory with a generic heavy field $\Phi$ and generic light fields; then, based on Ref. \cite{Henning:2014wua}, we can write the relevant part of the Lagrangian as:
\be
\label{eq:the-toy-Lag}
\mathcal{L} \ni \Big\{ \Phi^{\dagger} B + \big[\Phi^{\dagger} V_1\big] \big[\Phi^{\dagger} V_2\big] + \mathrm{h.c.} \Big\} + \Phi^{\dagger} \Big(-D^2 - M_2 - U \Big) \Phi,
\ee
where $B, V_1, V_2$ and $U$ are generic functions of the light fields, $D^2=D^{\mu}D_{\mu}$, $M_2$ is a real parameter (representing the squared mass of $\Phi$) and we neglect terms that are cubic or higher in $\Phi$.%
\fn{In the 2HDM model of interest here, there are no cubic scalar terms consistent with the gauge symmetry.}
The equation of motion for $\Phi$ is thus:
\be
\label{eq:EoM-Phi}
(-D^2 - M_2 - U) \Phi - \big[\Phi^{\dagger} V_1\big] V_2 - \big[\Phi^{\dagger} V_2\big] V_1 + B = 0.
\ee
Since $\Phi$ is heavy, we assume that the solution for $\Phi$ can be written as
\be
\label{eq:Hermes-ansatz}
\Phi_c = \Phi_c^{(0)} + \dfrac{\Phi_c^{(1)}}{M_2} + \dfrac{\Phi_c^{(2)}}{{M_2}^2} + ... \, .
\ee
The different values $\Phi_c^{(i)}$ can be determined by inserting Eq. \ref{eq:Hermes-ansatz} in Eq. \ref{eq:EoM-Phi} and solving order by order in ${M_2}^{-1}$. After this, the effective Lagrangian can be obtained by replacing $\Phi$ in Eq. \ref{eq:the-toy-Lag} by $\Phi_c$, up to the desired order. 
Now, if we intend to truncate the effective Lagrangian with ${M_2}^{-2}$ terms, we do not need $\Phi_c^{(2)}$ or higher orders. To see this, we start by noting that
\be
\Phi_c^{(0)} =0, 
\hspace{15mm}
\Phi_c^{(1)} =B, 
\ee
which can be trivially obtained by solving Eq. \ref{eq:EoM-Phi} to order ${M_2}^{1}$ and ${M_2}^0$, respectively. Using these results, and inserting the solution Eq.  \ref{eq:Hermes-ansatz} (up to the $\Phi_c^{(2)}$ term) in Eq. \ref{eq:the-toy-Lag}, the terms up to order ${M_2}^{-2}$ are
\bea
\mathcal{L} &\ni&
\Bigg\{ \Bigg[\dfrac{B}{M_2} + \dfrac{\Phi_c^{(2)}}{{M_2}^2}\Bigg]^{\dagger} B + \Bigg[\dfrac{B}{M_2} + \dfrac{\Phi_c^{(2)}}{{M_2}^2}\Bigg]^{\dagger} V_1 \Bigg[\dfrac{B}{M_2} + \dfrac{\Phi_c^{(2)}}{{M_2}^2}\Bigg]^{\dagger} V_2 + \mathrm{h.c.} \Bigg\} \nonumber \\
&& \qquad \qquad + \Bigg[\dfrac{B}{M_2} + \dfrac{\Phi_c^{(2)}}{{M_2}^2}\Bigg]^{\dagger} \Big(-D^2 - M_2 - U \Big) \Bigg[\dfrac{B}{M_2} + \dfrac{\Phi_c^{(2)}}{{M_2}^2}\Bigg], \nonumber\\[3mm]
&=& \dfrac{|B^2|}{M_2} + \left\{ \dfrac{(B^{\dagger} V_1) (B^{\dagger} V_2)}{{M_2}^2} + \mathrm{h.c.} \right\} + \dfrac{B^{\dagger} (D^2 - U) B}{{M_2}^2} + \mathcal{O}\left(\dfrac{1}{{M_2}^3}\right),
\label{eq:cde}
\eea
so that the $\Phi_c^{(2)}$ contribution cancels, as predicted.
Note that we have assumed that there is only one heavy mass scale, $M_2$; if there were multiple  heavy scales, there would be additional contributions.

We now apply this generic procedure to the particular case of the 2HDM.
It should be clear from the start that, in order for the EFT for the 2HDM to be valid, the heavy degrees of freedom of the 2HDM must decouple, which requires that the heavy masses defined implicitly in Eqs. \ref{eq:zmass1} to \ref{eq:zmass} all be much heavier than the weak scale, $m_A^2 \sim m_{H_0}^2 \sim m_{H^{\pm}}^2 \gg v^2$. Accordingly, we assume one heavy scale $Y_2 \equiv \Lambda^2$, obeying $Y_2  \gg v^2$. Following the procedure defined above, then, we can write:
\be
\label{eq:2HDM-sol}
{H_2}_c = {H_2}_c^{(0)} + \dfrac{{H_2}_c^{(1)}}{Y_2} + \dfrac{{H_2}_c^{(2)}}{{Y_2}^2} + ... \, .
\ee
As in the generic case, we  conclude that ${H_2}_c^{(0)} = 0$ and:%
\footnote{Although we will only present numerical results for the scenario where the parameters $Y_3, Z_{5}, Z_{6}, Z_{7}$ take real values, we consider them in this section as generally complex.
Besides, from here on, we omit the lepton Yukawa couplings for pedagogical clarity; they follow the contributions from the down-type quark Yukawa couplings in an obvious fashion.}
\be
{H_2}_c^{(1)}
=
- Y_3^* H_{1}
-
Z_{6}^* \left(H_{1}^{\dagger} H_{1}\right) H_{1}
-
\lambda_u^{(2)*} \widehat{q}_L u_R
-
\lambda_d^{(2)} \bar{d}_R q_L.
\ee

Again as in the generic case, we can neglect ${H_2}_c^{(2)}$ and higher orders, as we want to truncate the effective Lagrangian with $Y_2^2$ terms. The effective Lagrangian is then obtained by replacing all occurences of $H_2$ in $\mathcal{L}_{\mathrm{2HDM}}$ by Eq. \ref{eq:2HDM-sol}. 
Now, following the terminology of Ref. \cite{Egana-Ugrinovic:2015vgy}, we identify terms with inverse powers of $Y_2$ of 0, 1 and 2 as having \textit{effective-dimension} (EFD) 4, 6 and 8, respectively, and label them $F_4$, $F_6$ and $F_8$. Then, we split the different $F_i$ (with $i$ being the EFD) into several $F_{i,j}$ terms, with $j$ being the \textit{absolute-dimension} (ABD).
Accordingly, the effective Lagrangian reads:
\be
\label{eq:4}
\mathcal{L}_{\mathrm{eff}} = F_{4} + \dfrac{F_{6}}{Y_2} + \dfrac{F_{8}}{Y_2^2}
+ \mathcal{O}\left(\dfrac{1}{Y_2^3}\right),
\ee
with:
\bs
\bea
\label{eq:6ap}
F_4 &=& F_{4,2} + F_{4,4} \, , \\
\label{eq:6bp}
F_{6} &=& 
F_{6,2} + F_{6,4} + F_{6,6} \, ,\\
\label{eq:6c}
F_8 &=&
F_{8,4} + F_{8,6} + F_{8,8} \, .
\eea
\es
The terms of EFD 4 are
\bs
\label{eq:6}
\bea
F_{4,2} &=& - Y_1 H_{1}^{\dagger} H_{1}, \\
\label{eq:6b}
F_{4,4} &=& \left(D_{\mu} H_1\right)^{\dagger} \left(D^{\mu} H_1\right) - \frac{Z_{1}}{2}\left(H_{1}^{\dagger} H_{1}\right)^{2} - \Big(
\lambda_u^{(1)} \bar{u}_R {\widehat{q}_L}^{\dagger} H_1
+
\lambda_d^{(1)} \bar{d}_R H_1^{\dagger} q_L
+ \textrm{h.c.} \Big),
\eea
\es
and the terms of EFD 6
\bs
\label{eq:7}
\bea
\label{eq:7a}
F_{6,2} &=& {|Y_3|}^2 (H_1^{\dagger} H_1), \\
\label{eq:7b}
F_{6,4} &=& 
Y_3 \lambda_u^{(2)*} H_1^{\dagger} \widehat{q}_L u_R
+
Y_3 \lambda_d^{(2)} \bar{d}_R H_1^{\dagger} q_L
+
Y_3 Z_{6}^* (H_1^{\dagger} H_1)^2
+
\mathrm{h.c.},\\
\label{eq:7c}
F_{6,6} &=& 
(H_1^{\dagger} H_1) 
\bigg[
|Z_{6}|^2
(H_1^{\dagger} H_1)^2
+
\Big\{
Z_{6} \lambda_u^{(2)*} H_1^{\dagger} \widehat{q}_L u_R
+
Z_{6} \lambda_d^{(2)}\bar{d}_R H_1^{\dagger} q_L
+
\mathrm{h.c.}
\Big\}
\bigg]
+
\mathrm{4F},
\eea
\es
where 4F represents operators involving four fermion fields; we consistently ignore them in the following, as they do not affect the lowest order Higgs couplings. 
Finally, for the terms of EFD 8, 
\bs
\label{eq:8}
\bea
\label{eq:8a}
F_{8,4} &=&
{|Y_3|}^2 (D_{\mu} H_1)^{\dagger} (D^{\mu} H_1)
-(H_1^{\dagger} H_1)^2
\left[
{|Y_3|}^2 Z_{34} + \dfrac{1}{2} (Y_3)^2 Z_{5}^* + \dfrac{1}{2} (Y_3^*)^2 Z_{5}
\right]
\, , \\
\label{eq:8b}
F_{8,6} &=&
\left\{Y_3 Z_{6}^* + Y_3^* Z_{6} \right\} (H_1^{\dagger} H_1) (D_{\mu} H_1)^{\dagger} (D^{\mu} H_1) 
+
\left\{ Y_3 Z_{6}^*
(D_{\mu} H_1)^{\dagger} H_1 + \mathrm{h.c.} \right\}
\partial^{\mu}(H_1^{\dagger} H_1) \no
&& + \bigg\{ 
Y_3^* \lambda_u^{(2)} \Big(D_{\mu} (\widehat{q}_L u_R) \Big)^{\dagger} (D^{\mu} H_1)
+
Y_3^* \lambda_d^{(2)*} \Big(D_{\mu} (\bar{d}_R q_L) \Big)^{\dagger} (D^{\mu} H_1)
+ \mathrm{h.c.} \bigg\} \no
&& -(H_1^{\dagger} H_1)^3 
\left[
Y_3 Z_{34} Z_{6}^* + 
Y_3 Z_{5}^* Z_{6} + \mathrm{h.c.}
\right] \no
&& - (H_1^{\dagger} H_1)
\bigg[
H_1^{\dagger} \widehat{q}_L u_R \Big(Y_3 Z_{34} \lambda_u^{(2)*} + Y_3^* Z_{5} \lambda_u^{(2)*}\Big)
+
\bar{d}_R H_1^{\dagger} q_L \Big(Y_3 Z_{34} \lambda_d^{(2)} + Y_3^* Z_{5} \lambda_d^{(2)}\Big)
+ \mathrm{h.c.} \bigg]
\, , \\
\label{eq:8c}
F_{8,8} &=&
{|Z_{6}|}^2 (H_1^{\dagger} H_1)^2 (D_{\mu} H_1)^{\dagger} (D^{\mu} H_1) 
+ 2 {|Z_{6}|}^2
(H_1^{\dagger} H_1)
\partial_{\mu}(H_1^{\dagger} H_1) \partial^{\mu}(H_1^{\dagger} H_1) \no
&& -(H_1^{\dagger} H_1)^4
\left[
Z_{34} {|Z_{6}|}^2 + \dfrac{1}{2} Z_{5}^* Z_{6}^2 + \dfrac{1}{2} Z_{5} (Z_{6}^*)^2\right] \no
&& - (H_1^{\dagger} H_1)^2 \bigg[
H_1^{\dagger} \widehat{q}_L u_R \Big(Z_{34} Z_{6} \lambda_u^{(2)*} + Z_{5} Z_{6}^* \lambda_u^{(2)*}\Big)
+
\bar{d}_R H_1^{\dagger} q_L \Big(Z_{34} Z_{6} \lambda_d^{(2)} + Z_{5} Z_{6}^* \lambda_d^{(2)}\Big) 
+ \mathrm{h.c.} \bigg] \no
&& +
\Bigg\{
\Big[
Z_{6}^* \lambda_u^{(2)} \Big(D_{\mu} (\widehat{q}_L u_R) \Big)^{\dagger}
+
Z_{6}^* \lambda_d^{(2)*} \Big(D_{\mu} (\bar{d}_R q_L) \Big)^{\dagger}
\Big]
\Big[
\partial^{\mu}(H_1^{\dagger} H_1) H_1
+
(H_1^{\dagger} H_1) (D^{\mu} H_1)
\Big]
+ \mathrm{h.c.}
\Bigg\} \no
&& 
+ \mathrm{4F},
\eea
\es
where $Z_{34} \equiv Z_{3} + Z_{4}$.

\subsection{Introduction to the conversion to SMEFT}

The theory described in Eqs. \ref{eq:4} to \ref{eq:8} is an EFT for the 2HDM, resulting from integrating out $H_2$. 
It contains an implicit \textit{matching} between the EFT and the 2HDM --- that is, a correspondence between the coefficient of an operator in the EFT and the parameters of the 2HDM.
We now want to write the theory in the \textit{SMEFT} format --- meaning, write it as the SM Lagrangian plus operators of ABD 6 (in the Warsaw basis\cite{Buchmuller:1985jz,Grzadkowski:2010es}) plus operators of ABD 8 (in the basis of Ref. \cite{Murphy:2020rsh}) ---, and render the matching explicit. To that end, we start by noting that the terms in SMEFT containing operators of ABD 2 and 4 are those of the SM, and only those;
to see this explicitly, the relevant terms in the SMEFT containing operators of ABD 2 and 4 can be written as follows:
\be
\label{eq:SM1}
\mathcal{L}_{\mathrm{SMEFT}} \ni S_{4,2} + S_{4,4},
\ee
where we are following the same convention for the subscripts, and where
\bs
\label{eq:SM2}
\bea
\label{eq:SM2-a}
S_{4,2} &=& \mu^2 \mathcal{H}^{\dagger} \mathcal{H}, \\
\label{eq:SM2-b}
S_{4,4} &=& \left(D_{\mu} \mathcal{H}\right)^{\dagger} \left(D^{\mu} \mathcal{H}\right) - \lambda \left(\mathcal{H}^{\dagger} \mathcal{H}\right)^{2} - \Big(
Y_u \, \bar{u}_R {\widehat{q}_L}^{\dagger} \mathcal{H}
+
Y_d \, \bar{d}_R \mathcal{H}^{\dagger} q_L
+ \textrm{h.c.} \Big),
\eea
\es
with $\mathcal{H}$ being the SMEFT Higgs doublet.
This implies that Eqs. \ref{eq:4} to \ref{eq:8} do not allow a direct comparison with the SMEFT, despite appearances. In fact, if one would naively identify $H_1$ with $\mathcal{H}$, the comparison would seem straightforward (given the similarity between $S_{4,2}$ and $F_{4,2}$, on the one hand, and between $S_{4,4}$ and $F_{4,4}$, on the other hand). So, for example, $Y_d$ would be identified with $\lambda_d^{(1)}$. A closer look, however, reveals that such an identification is not correct, as there are other terms with the operator $\bar{d}_R H_1^{\dagger} q_L$ in Eqs. \ref{eq:7} and \ref{eq:8} besides the one in $F_{4,4}$. Therefore, in order for the EFT of Section \ref{sec:IOH2} to be interpreted in the SMEFT environment, a careful identification needs to be done.
Such  an identification is rendered even less direct due to the need for basis conversions; indeed, as mentioned in passing, we want to write all the operators of ABD 6 in the Warsaw basis and all those of ABD 8 in the basis of Ref. \cite{Murphy:2020rsh}.%
\fn{Although the basis of Ref. \cite{Murphy:2020rsh} is equivalent to that of Ref. \cite{Li:2020xlh}, we stick to the notation of the former.}
This conversion will generate operators of ABD 4, which need to be taken into account for a proper identification.

In what follows, we perform the basis conversions; we identify $H_1$ as $H$, to simplify the notation.

\subsection{Operators of ABD 6}
We begin with the terms with ABD 6 operators; only those in the first two lines of $F_{8,6}$ have operators which do not belong to the Warsaw basis. Let us start with the second term of the first line; 
\bea
\label{eq:10}
&& \left\{ Y_3 Z_{6}^*
(D_{\mu} H)^{\dagger} H + \mathrm{h.c.} \right\}
\partial^{\mu}(H^{\dagger} H)
=
\left\{ Y_3 Z_{6}^*
(D_{\mu} H)^{\dagger} H + \mathrm{h.c.} \right\}
D^{\mu}(H^{\dagger} H) \no
= &&
- (H^{\dagger} H) \left\{ Y_3 Z_{6}^* D^{\mu} \Big[ (D_{\mu} H)^{\dagger} H \Big] + \mathrm{h.c.} \right\} \no
= && - Y_3 Z_{6}^* (H^{\dagger} H) \Big[ (D^2 H)^{\dagger} H + (D_{\mu} H)^{\dagger} (D^{\mu} H)\Big] + \mathrm{h.c.}\, .
\label{eq:reds}
\eea
We see that the term with $(D_{\mu} H)^{\dagger} (D^{\mu} H)$ in Eq. \ref{eq:reds} precisely cancels the first term of $F_{8,6}$. 
As for the term with $D^2 H$, we can use the Equation of Motion (EoM) for $H$,%
\fn{When including operators beyond dimension 6, one should generally use field redefinitions to match onto a given basis of dimension-6 operators\cite{Criado:2018sdb}. In this particular case, though, the operators that we are rewriting are EFD 8. Thus, the difference between using field redefinitions and using equations of motion will be $\mathcal{O}(\Lambda^{-6})$, which is beyond the order to which we are matching.}
\be
\label{eq:EoM}
D^2 H = -Y_1 H - Z_{1} (H^{\dagger} H) H - \lambda_u^{(1)} \widehat{q}_L u_R
-\lambda_d^{(1)} \bar{d}_R q_L
.
\ee
Inserting this equality in the term with $D^2 H$ in Eq. \ref{eq:10} gives:
\be
Y_3 Z_{6}^* \Bigg[ Y_1 (H^{\dagger} H)^2 + Z_{1} (H^{\dagger} H)^3
+ \lambda_u^{(1)} (H^{\dagger} H) \bar{u}_R \widehat{q}_L^{\dagger} H
+ \lambda_d^{(1)} (H^{\dagger} H) \bar{q}_L H d_R \Bigg]
+ \mathrm{h.c.} \, .
\ee
Here, the operator of the first term is ABD 4, while
the operators in the other three terms which have ABD 6 belong to the Warsaw basis.
Finally, for the two terms in the second line of $F_{8,6}$, we can use Integration by Parts (IbP) and Eq. \ref{eq:EoM} to conclude that they are respectively given by:
\bs
\bea
&& Y_3^* \lambda_u^{(2)} \Bigg[ Y_1 \bar{u}_R \widehat{q}_L^{\dagger} H + Z_{1} (H^{\dagger} H) \bar{u}_R \widehat{q}_L^{\dagger} H + \mathrm{4F} \Bigg] + \mathrm{h.c.}, \\
&& Y_3^* \lambda_d^{(2)*} \Bigg[ Y_1 \bar{q}_L H d_R + Z_{1} (H^{\dagger} H) \bar{q}_L H d_R + \mathrm{4F}\Bigg] + \mathrm{h.c.} \, .
\eea
\es
As before, the operators in the first terms have ABD 4, whereas the other operators belong to the Warsaw basis.

\subsection{Operators of ABD 8}
We now consider the terms with operators of ABD 8;
using IbP and the EoM, we find that the second term of $F_{8,8}$ can be rewritten as
\bea
&& 2 {|Z_{6}|}^2
(H^{\dagger} H)
\partial_{\mu}(H^{\dagger} H) \partial^{\mu}(H^{\dagger} H)
=
2 Y_1 {|Z_{6}|}^2 (H^{\dagger} H)^3
+ {|Z_{6}|}^2 \Big\{ \lambda_u^{(1)} (H^{\dagger} H)^2 \bar{q}_L u_R \tilde{H} \no
&& \qquad + \lambda_d^{(1)} (H^{\dagger} H)^2 \bar{d}_R H^{\dagger} q_L
+ \mathrm{h.c.} \Big\}
+ 2 Z_1 {|Z_{6}|}^2 (H^{\dagger} H)^4 
- 2 {|Z_{6}|}^2 (H^{\dagger} H)^2 (D_{\mu} H_1)^{\dagger} (D^{\mu} H_1)\, ,
\label{eq:more} 
\eea
where the first term on the right of Eq. \ref{eq:more} has ABD 6.
As for the terms inside the curly brackets in the third line of $F_{8,8}$, we can use SU(2) identities to see that:%
\fn{Here, $\tau^I$ are the Pauli matrices. Besides, as in the case of $\tilde{H}_i$, we have $\widetilde{D^{\mu} H} = i \sigma_2 D^{\mu} H$.}
\bs
\bea
\Big[(D_{\mu} H)^{\dagger} \tau^I (D^{\mu} H) \Big] \Big[\bar{q}_L u_R \tau^I \tilde{H} \Big]
=
\Big[(D_{\mu} H)^{\dagger} (D^{\mu} H) \Big] \Big[\bar{q}_L u_R \tilde{H}\Big]
- 2 \Big[H^{\dagger} (D_{\mu} H)\Big] \Big[\bar{q}_L u_R \widetilde{D^{\mu} H}\Big], \\
\Big[(D_{\mu} H)^{\dagger} \tau^I (D^{\mu} H) \Big] \Big[\bar{q}_L d_R \tau^I H\Big]
=
2 \Big[(D_{\mu} H)^{\dagger} H \Big] \Big[\bar{q}_L d_R D^{\mu} H\Big] - \Big[(D_{\mu} H)^{\dagger} (D^{\mu} H) \Big] \Big[\bar{q}_L d_R H\Big],
\eea
\es
which, together with IbP and the EoM, yields
\bs
\bea
&& Z_{6}^* \lambda_u^{(2)} \Big(D_{\mu} (\widehat{q}_L u_R) \Big)^{\dagger} 
\Big[
\partial^{\mu}(H_1^{\dagger} H_1) H_1
+
(H_1^{\dagger} H_1) (D^{\mu} H_1)
\Big]
+ \mathrm{h.c.} \no
&& = Z_{6} \lambda_u^{(2)*} \bigg\{
3 Y_1 (H^{\dagger} H) \bar{q}_L u_R \tilde{H}
+ 3 Z_1 (H^{\dagger} H)^2 \bar{q}_L u_R \tilde{H}
- 3 (D_{\mu} H)^{\dagger} (D^{\mu} H) \bar{q}_L u_R \tilde{H} \no
&& \qquad + \Big[(D_{\mu} H)^{\dagger} \tau^I (D^{\mu} H) \Big] \Big[\bar{q}_L u_R \tau^I \tilde{H} \Big]
- 2 \Big[(D_{\mu}H)^{\dagger} H \Big] \Big[ \bar{q}_L u_R \widetilde{D^{\mu} H} \Big] + \mathrm{4F} \bigg\} + \mathrm{h.c.} \, , \\[3mm]
&& Z_{6}^* \lambda_d^{(2)*} \Big(D_{\mu} (\bar{d}_R \, q_L) \Big)^{\dagger} 
\Big[
\partial^{\mu}(H_1^{\dagger} H_1) H_1
+
(H_1^{\dagger} H_1) (D^{\mu} H_1)
\Big]
+ \mathrm{h.c.} \no
&& = Z_{6}^* \lambda_d^{(2)*} \bigg\{
3 Y_1 (H^{\dagger} H) \bar{q}_L d_R H
+ 3 Z_1 (H^{\dagger} H)^2 \bar{q}_L d_R H
- 3 (D_{\mu} H)^{\dagger} (D^{\mu} H) \bar{q}_L d_R H \no
&& \qquad - \Big[(D_{\mu} H)^{\dagger} \tau^I (D^{\mu} H) \Big] \Big[\bar{q}_L d_R \tau^I H\Big]
- 2 (H^{\dagger} D_{\mu} H) \bar{q}_L d_R D^{\mu} H + \mathrm{4F} \bigg\} + \mathrm{h.c.} \, .
\eea
\es

\subsection{Results}
Putting the pieces together, we have
\be
\label{eq:the-res-Lag}
\mathcal{L}_{\mathrm{eff}} = F_{4} + \dfrac{F_{6}}{Y_2} + \dfrac{F'_{8}}{Y_2^2}
+ \mathcal{O}\left(\dfrac{1}{Y_2^3}\right),
\ee
where, for the EFD 4 and 6 terms, we have the same result as previously (Eqs. \ref{eq:6} and \ref{eq:7}, respectively), while for the EFD 8 contribution we have
\be
\label{eq:6c-new}
F'_8 = F'_{8,4} + F'_{8,6} + F'_{8,8} \, ,
\ee
with:
\bs
\label{eq:8prime}
\bea
\label{eq:8aprime}
F'_{8,4} &=&
{|Y_3|}^2 (D_{\mu} H)^{\dagger} (D^{\mu} H)
-(H^{\dagger} H)^2
\Big[
{|Y_3|}^2 Z_{34} + \dfrac{1}{2} (Y_3)^2 Z_{5}^* + \dfrac{1}{2} (Y_3^*)^2 Z_{5} - Y_1 Y_3 Z_6^* \no
&& - Y_1 Y_3^* Z_6 \Big] + \Big \{Y_1 Y_3 \lambda_u^{(2)*} H^{\dagger} \widehat{q}_L u_R
+ Y_1 Y_3 \lambda_d^{(2)} \bar{d}_R H^{\dagger} q_L 
+ \mathrm{h.c.} \Big\}
\, , \\
\label{eq:8bprime}
F'_{8,6} &=&
-(H^{\dagger} H)^3 
\left[
Y_3 Z_{34} Z_{6}^* + 
Y_3 Z_{5}^* Z_{6} -
Y_3 Z_1 Z_6^* -
Y_1 |Z_6|^2
+ \mathrm{h.c.}
\right] \no
&& - (H^{\dagger} H)
\bigg\{
H^{\dagger} \widehat{q}_L u_R \Big(Y_3 Z_{34} \lambda_u^{(2)*} + Y_3^* Z_{5} \lambda_u^{(2)*}
- Y_3^* Z_6 \lambda_u^{(1)} - Y_3 Z_1 \lambda_u^{(2)*} - 3 Y_1 Z_6 \lambda_u^{(2)*} \Big) \no
&& \hspace{13mm} +
\bar{d}_R H^{\dagger} q_L \Big(Y_3 Z_{34} \lambda_d^{(2)} + Y_3^* Z_{5} \lambda_d^{(2)}
- Y_3^* Z_6 \lambda_d^{(1)} - Y_3 Z_1 \lambda_d^{(2)} - 3 Y_1 Z_6 \lambda_d^{(2)} \Big)
+ \mathrm{h.c.} \bigg\} \no
&& + ~\mathrm{4F}
\, , \\
\label{eq:8cprime}
F'_{8,8} &=&
-{|Z_{6}|}^2 (H^{\dagger} H)^2 (D_{\mu} H)^{\dagger} (D^{\mu} H) 
\no
&& -(H^{\dagger} H)^4
\left[
Z_{34} {|Z_{6}|}^2 + \dfrac{1}{2} Z_{5}^* Z_{6}^2 + \dfrac{1}{2} Z_{5} (Z_{6}^*)^2 - 2 Z_1 |Z_6|^2 \right] \no
&& - (H^{\dagger} H)^2 \bigg\{
H^{\dagger} \widehat{q}_L u_R \Big(Z_{34} Z_{6} \lambda_u^{(2)*} + Z_{5} Z_{6}^* \lambda_u^{(2)*} - |Z_6|^2 \lambda_u^{(1)} - 3 Z_1 Z_6 \lambda_u^{(2)*}\Big) \no
&& \hspace{15mm} +
\bar{d}_R H^{\dagger} q_L \Big(Z_{34} Z_{6} \lambda_d^{(2)} + Z_{5} Z_{6}^* \lambda_d^{(2)} - |Z_6|^2 \lambda_d^{(1)} - 3 Z_1 Z_6 \lambda_d^{(2)} \Big) 
+ \mathrm{h.c.} \bigg\} \no
&&
- 3 (D_{\mu} H)^{\dagger} (D^{\mu} H) \Big\{Z_6 \lambda_u^{(2)*} H^{\dagger} \widehat{q}_L u_R + Z_6^* \lambda_d^{(2)*} \bar{q}_L d_R H + \mathrm{h.c.} \Big\}
\no
&&
+ \Bigg\{ Z_{6} \lambda_u^{(2)*} \bigg(\Big[(D_{\mu} H)^{\dagger} \tau^I (D^{\mu} H) \Big] \Big[\bar{q}_L u_R \tau^I \tilde{H} \Big]
- 2 \Big[(D_{\mu}H)^{\dagger} H \Big] \Big[ \bar{q}_L u_R \widetilde{D^{\mu} H} \Big] \bigg)
\no
&& 
\hspace{3mm} - Z_{6}^* \lambda_d^{(2)*} \bigg(\Big[(D_{\mu} H)^{\dagger} \tau^I (D^{\mu} H) \Big] \Big[\bar{q}_L d_R \tau^I H\Big]
+ 2 (H^{\dagger} D_{\mu} H) \bar{q}_L d_R D^{\mu} H \bigg) + \mathrm{h.c.} \Bigg\}
\no
&& 
+ \mathrm{4F}.
\eea
\es

\subsection{The SMEFT Lagrangian}
We are now ready to write the effective Lagrangian in the SMEFT formalism, which involves determining the matching coefficients up to EFD 8.
In particular, we can determine all the coefficients involved in Eq. \ref{eq:SM1};
for example, by equating all the terms of ABD 2 of Eq. \ref{eq:the-res-Lag} (namely, $F_{4,2}$ and $F_{6,2}$) with all the terms of ABD 2 of Eq. \ref{eq:SM1} (namely, $S_{4,2}$), we find:
\be
\label{eq:irrelevant}
\mu^2 = - Y_1 + \dfrac{|Y_3|^2}{Y_2}.
\ee
A similar exercise can be done for the remaining parameters of Eq. \ref{eq:SM1}. This exercise is nothing but the matching between SMEFT and the 2HDM for the operators of ABD 2 and 4. From the point of view of the SMEFT, however, only the matching for the operators of ABD greater than 4 is relevant; the reason is that relations like Eq. \ref{eq:irrelevant} correspond to redefinitions of SM parameters, so that they have no observable effect beyond the SM.
On the other hand, care should be taken with the kinetic terms of the Higgs doublet; in fact, in order to obtain a proper correspondence between Eqs. \ref{eq:SM1} and \ref{eq:the-res-Lag} for the operators of ABD 4, we must have:
\be
\left(D_{\mu} \mathcal{H}\right)^{\dagger} \left(D^{\mu} \mathcal{H}\right)
=
\left(D_{\mu} H\right)^{\dagger} \left(D^{\mu} H\right) \Bigg(1+ \dfrac{{|Y_3|}^2}{Y_2^2}\Bigg),
\ee
which leads to,
\be
H = \mathcal{H} \Bigg(1 - \dfrac{{|Y_3|}^2}{2 Y_2^2}\Bigg).
\ee
This relation is the last element required to write the SMEFT Lagrangian and perform the matching of operators of ABD 6 and 8; replacing $Y_2$ by the heavy (squared) scale $\Lambda^2$, we write it as:
\be
\label{eq:SMEFT-Lag}
\mathcal{L}_{\mathrm{SMEFT}} = \mathcal{L}_{\mathrm{SM}} + \dfrac{S_{\text{all},6}}{\Lambda^2} + \dfrac{S_{\text{all},8}}{\Lambda^4} + \mathcal{O}\left(\dfrac{1}{\Lambda^6}\right),
\ee
where the subscript `all' means that  all EFDs are being included; moreover, $\mathcal{L}_{\mathrm{SM}}$ is given by Eqs. \ref{eq:SM1} and \ref{eq:SM2}, and:
\be
\label{eq:Sall6}
S_{\text{all},6} = 
C_{\mathcal{H}} (\mathcal{H}^{\dagger} \mathcal{H})^3
+ 
\left\{
C_{u\mathcal{H}} \, (\mathcal{H}^{\dagger} \mathcal{H}) \, \bar{q}_L u_R \tilde{\mathcal{H}} +
C_{d\mathcal{H}} \, (\mathcal{H}^{\dagger} \mathcal{H}) \, \bar{q}_L d_R \mathcal{H} +
\mathrm{h.c.} \right\} + \mathrm{4F},
\ee
with:
\bs
\label{eq:matching:ABD6}
\bea
C_{\mathcal{H}} \hspace{-2mm} &=& \hspace{-2mm} C_\mathcal{H}^{[6]} + \dfrac{C_\mathcal{H}^{[8]}}{\Lambda^2} \no
&=& \hspace{-2mm} |Z_{6}|^2 + \dfrac{1}{\Lambda^2} \Big( Y_3 Z_{1} Z_{6}^* + Y_3^* Z_{1} Z_{6}
- Y_3 Z_{34} Z_{6}^* - Y_3^* Z_{34} Z_{6}
- Y_3 Z_{5}^* Z_{6} - Y_3^* Z_{5} Z_{6}^* + 2 Y_1 {|Z_{6}|}^2 \Big), \hspace{5mm} \\[3mm]
C_{u\mathcal{H}} \hspace{-2mm} &=& \hspace{-2mm} C_{u\mathcal{H}}^{[6]} + \dfrac{C_{u\mathcal{H}}^{[8]}}{\Lambda^2} \no
&=& \hspace{-2mm} Z_{6} \lambda_u^{(2)*} + \dfrac{1}{\Lambda^2} \Big( Y_3^* Z_{6} \lambda_u^{(1)}
+ Y_3 Z_{1} \lambda_u^{(2)*} - Y_3 Z_{34} \lambda_u^{(2)*} - Y_3^* Z_{5} \lambda_u^{(2)*} + 3 Y_1 Z_{6} \lambda_u^{(2)*} \Big), \\[3mm]
C_{d\mathcal{H}} \hspace{-2mm} &=& \hspace{-2mm} C_{d\mathcal{H}}^{[6]} + \dfrac{C_{d\mathcal{H}}^{[8]}}{\Lambda^2} \no
&=& \hspace{-2mm} Z_{6}^* \lambda_d^{(2)*} + \dfrac{1}{\Lambda^2} \Big( Y_3 Z_{6}^* \lambda_d^{(1)} + Y_3^* Z_{1} \lambda_d^{(2)*} - Y_3^* Z_{34} \lambda_d^{(2)*} - Y_3 Z_{5}^* \lambda_d^{(2)*} + 3 Y_1 Z_{6}^* \lambda_d^{(2)*} \Big),
\eea
\es
where the superscript $[i]$ indicates that the EFD at stake is $i$, and
\bea
\label{eq:Sall8}
S_{\text{all},8} &=&
C_{\mathcal{H}^8} (\mathcal{H}^{\dagger} \mathcal{H})^4
+
C_{\mathcal{H}^6}^{(1)} (\mathcal{H}^{\dagger} \mathcal{H})^2 \left(D_{\mu} \mathcal{H}\right)^{\dagger} \left(D^{\mu} \mathcal{H}\right)
+
\Big\{
C_{qu\mathcal{H}^5} (\mathcal{H}^{\dagger} \mathcal{H})^2 \bar{q}_L u_R \tilde{\mathcal{H}} \no
&& +
C_{qu\mathcal{H}^3D^2}^{(1)} (D_{\mu} \mathcal{H})^{\dagger} (D^{\mu} \mathcal{H}) \bar{q}_L u_R \tilde{\mathcal{H}}
+
C_{qu\mathcal{H}^3D^2}^{(2)} \Big[(D_{\mu} \mathcal{H})^{\dagger} \tau^I (D^{\mu} \mathcal{H}) \Big] \Big[\bar{q}_L u_R \tau^I \tilde{\mathcal{H}}\Big] \no
&& 
+
C_{qu\mathcal{H}^3D^2}^{(5)} \Big[(D_{\mu}\mathcal{H})^{\dagger} \mathcal{H} \Big] \Big[ \bar{q}_L u_R \widetilde{D^{\mu} \mathcal{H}} \Big]
+
C_{qd\mathcal{H}^5} (\mathcal{H}^{\dagger} \mathcal{H})^2 \bar{q}_L d_R \mathcal{H} \no
&& +
C_{qd\mathcal{H}^3D^2}^{(1)} (D_{\mu} \mathcal{H})^{\dagger} (D^{\mu} \mathcal{H}) \bar{q}_L d_R \mathcal{H}
+
C_{qd\mathcal{H}^3D^2}^{(2)} \Big[(D_{\mu} \mathcal{H})^{\dagger} \tau^I (D^{\mu} \mathcal{H}) \Big] \Big[\bar{q}_L d_R \tau^I \mathcal{H}\Big] \no
&& 
+
C_{qd\mathcal{H}^3D^2}^{(5)} (\mathcal{H}^{\dagger} D_{\mu} \mathcal{H}) (\bar{q}_L d_R D^{\mu} \mathcal{H})
+ \mathrm{h.c.} \Big\} + \mathrm{4F},
\eea
where:
\bs
\label{eq:matching:ABD8}
\bea
C_{\mathcal{H}^8} &=& - Z_{34} {|Z_{6}|}^2 - \dfrac{1}{2} Z_{5}^* Z_{6}^2 - \dfrac{1}{2} Z_{5} (Z_{6}^*)^2 + 2 Z_1 {|Z_{6}|}^2, \\
C_{\mathcal{H}^6}^{(1)} &=& - {|Z_{6}|}^2, \\
C_{qu\mathcal{H}^5} &=& - Z_{34} Z_{6} \lambda_u^{(2)*} - Z_{5} Z_{6}^* \lambda_u^{(2)*} + {|Z_{6}|}^2 \lambda_u^{(1)} + 3 Z_1 Z_{6} \lambda_u^{(2)*}, \\
C_{qu\mathcal{H}^3D^2}^{(1)} &=& - 3 Z_{6} \lambda_u^{(2)*}, \\
C_{qu\mathcal{H}^3D^2}^{(2)} &=& Z_{6} \lambda_u^{(2)*}, \\
C_{qu\mathcal{H}^3D^2}^{(5)} &=& - 2 Z_{6} \lambda_u^{(2)*}, \\
C_{qd\mathcal{H}^5} &=& - Z_{34} Z_{6}^* \lambda_d^{(2)*} - Z_{5}^* Z_{6} \lambda_d^{(2)*} + {|Z_{6}|}^2 \lambda_d^{(1)} + 3 Z_1 Z_{6}^* \lambda_d^{(2)*}, \\
C_{qd\mathcal{H}^3D^2}^{(1)} &=& - 3 Z_{6}^* \lambda_d^{(2)*}, \\
C_{qd\mathcal{H}^3D^2}^{(2)} &=& - Z_{6}^* \lambda_d^{(2)*}, \\
C_{qd\mathcal{H}^3D^2}^{(5)} &=& - 2 Z_{6}^* \lambda_d^{(2)*}.
\eea
\es

\subsection{The decoupling limit}
\label{sec:decoupling}

As already suggested, in order for the EFT for the 2HDM to be valid (either in the form of $\mathcal{L}_{\mathrm{eff}}$ of Eq. \ref{eq:the-res-Lag} or in the  form of $\mathcal{L}_{\mathrm{SMEFT}}$ of Eq. \ref{eq:SMEFT-Lag}), the heavy degrees of freedom of the 2HDM must decouple.
In the scenario where CP is conserved in the scalar sector, we characterize the \textit{decoupling limit}  \cite{Gunion:2002zf} by the assumptions:
\be
\label{eq:decoupling-real}
m_A^2 \sim m_{H_0}^2 \sim m_{H^{\pm}}^2
 \sim Y_2 \equiv \Lambda^2 \gg v^2,
\qquad
m_h^2 \simeq v^2,
\ee
as well as the assumption of perturbative unitarity. The latter is ensured by requiring $Z_i/(4 \pi) \simeq \mathcal{O}(1)$ \cite{Ferreira:2014naa}. To ensure this condition, in turn, Eqs. \ref{eq:theZs-real} can be used to conclude that the following relation must hold:
\be
\label{eq:camb-size}
|\cos (\beta-\alpha)| \propto \dfrac{v^2}{\Lambda^2}.
\ee
The decoupling limit thus implies the \textit{alignment limit}, which is characterized by $\cos (\beta-\alpha) \to 0$ and corresponds to the SM prediction.%
\fn{Using Eqs. \ref{eq:theYs}, \ref{eq:theZs-real}, it is easy to see that all the terms with operators of EFD 6 and 8 vanish in the limit $\cos (\beta-\alpha) \to 0$.} 
The relation \ref{eq:camb-size} means that, since we are working up EFD 8 (i.e. up to quadratic order in the expansion in powers of $1/\Lambda^2$), we can always expand the results to \textit{quadratic} order in $\cos (\beta-\alpha)$. Performing this expansion, we find:
\bs
\label{eq:WCs-after-decoupling}
\bea
\label{eq:CuH}
{ C_{u\mathcal{H}}\over \Lambda^2}
&=& 
- \dfrac{2}{\sqrt{2}} \, (\sqrt{2} G_F)^{3/2} \,  \dfrac{\cos (\beta-\alpha)}{\tan\beta}  \, \eta_u \, {m_u} \, 
\nonumber \\ && \qquad
- \dfrac{\cos (\beta-\alpha) \, {m_u} \, (\sqrt{2} G_F)^{3/2}}{\sqrt{2} \, {\Lambda}^2}  
\bigg[ \cos (\beta-\alpha) \, {\Lambda}^2 - \dfrac{6 \, m_h^2 \, \eta_u}{\tan\beta} \bigg], \\
{ C_{d\mathcal{H}}\over \Lambda^2}
&=&
- \dfrac{2}{\sqrt{2}} \, (\sqrt{2} G_F)^{3/2} \,  \dfrac{\cos (\beta-\alpha)}{\tan\beta}  \, \eta_d \, {m_d} \, 
\nonumber \\ && \qquad
- \dfrac{\cos (\beta-\alpha) \, {m_d} \, (\sqrt{2} G_F)^{3/2}}{\sqrt{2} \, {\Lambda}^2}  
\bigg[ \cos (\beta-\alpha) \, {\Lambda}^2 - \dfrac{6 \, m_h^2 \, \eta_d}{\tan\beta} \bigg], \\
\label{eq:cH}
{C_{\mathcal{H}}\over \Lambda^2}
&=&
\cos (\beta-\alpha)^2 \,  (\sqrt{2} G_F)^2  \left( {\Lambda}^2 - 4 \, m_h^2 \right)
\eea
\es
and
\bs
\label{eq:more-WCs-after-decoupling}
\bea
{C_{\mathcal{H}^8} \over \Lambda^4} &=& 2 \, \cos (\beta-\alpha)^2 \, m_h^2 \, (\sqrt{2} G_F)^3 
, \\
\label{eq:46b}
{C_{\mathcal{H}^6}^{(1)} \over \Lambda^4} &=& - \cos (\beta-\alpha)^2 \, (\sqrt{2} G_F)^2 
, \\
{C_{qu\mathcal{H}^5} \over \Lambda^4} &=& \dfrac{\sqrt{2} \, \cos (\beta-\alpha) \, {m_u} \, (\sqrt{2} G_F)^{5/2}}{{\Lambda}^2 }  \,  \bigg[\cos (\beta-\alpha) \, {\Lambda}^2 - \dfrac{3 \, m_h^2 \, \eta_u}{\tan\beta} \bigg]
, \\
{C_{qu\mathcal{H}^3D^2}^{(1)} \over \Lambda^4} &=& \dfrac{3 \, \sqrt{2} \, \cos (\beta-\alpha) \, {m_u} \, (\sqrt{2} G_F)^{3/2}  \, \eta_u}{{\tan \beta \, \Lambda}^2}
, \\
{C_{qu\mathcal{H}^3D^2}^{(2)} \over \Lambda^4} &=& - \dfrac{\sqrt{2} \, \cos (\beta-\alpha) \, {m_u} \, (\sqrt{2} G_F)^{3/2}  \, \eta_u}{{\tan \beta \, \Lambda}^2 }
, \\
{C_{qu\mathcal{H}^3D^2}^{(5)} \over \Lambda^4} &=& \dfrac{2 \, \sqrt{2} \, \cos (\beta-\alpha) \, {m_u} \, (\sqrt{2} G_F)^{3/2}  \, \eta_u}{{\tan \beta \, \Lambda}^2 }
, \\
{C_{qd\mathcal{H}^5} \over \Lambda^4} &=& \dfrac{\sqrt{2} \, \cos (\beta-\alpha) \, {m_d} \, (\sqrt{2} G_F)^{5/2}}{{\Lambda}^2 }  \,  \bigg[ \cos (\beta-\alpha) \, {\Lambda}^2 - \dfrac{3 \, m_h^2 \, \eta_d}{\tan\beta} \bigg]
, \\
{C_{qd\mathcal{H}^3D^2}^{(1)} \over \Lambda^4} &=& \dfrac{3 \, \sqrt{2} \, \cos (\beta-\alpha) \, {m_d} \, (\sqrt{2} G_F)^{3/2}  \, \eta_d}{{\tan \beta \, \Lambda}^2 }
, \\
{C_{qd\mathcal{H}^3D^2}^{(2)} \over \Lambda^4} &=& \dfrac{\sqrt{2} \, \cos (\beta-\alpha) \, {m_d} \, (\sqrt{2} G_F)^{3/2}  \, \eta_d}{{\tan \beta \, \Lambda}^2 }
, \\
{C_{qd\mathcal{H}^3D^2}^{(5)} \over \Lambda^4} &=& \dfrac{2 \, \sqrt{2} \, \cos (\beta-\alpha) \, {m_d} \, (\sqrt{2} G_F)^{3/2}  \, \eta_d}{{\tan \beta \, \Lambda}^2 },
\eea
\es
where, as before, $m_u$ and $m_d$ are a compact way to accomodate the three generations of masses of up-type and down-type quarks, respectively.%
\fn{We continue to  omit the terms for the leptons, which are trivially equivalent to the ones for the down-type quarks.}

\subsection{Relations in SMEFT}
\label{sec:relations}

Finally, in order to derive results in SMEFT, we need to consider two aspects. The first one involves the SMEFT Higgs doublet, which can be parametrized as
\be
\label{eq:SM-doublet}
\mathcal{H}
= 
\begin{pmatrix}
G^+_{\text{\tiny S}} \\
\frac{1}{\sqrt{2}}(v_T + h_{\text{\tiny S}} + i G_{0,\text{\tiny S}})
\end{pmatrix},
\ee
where $v_T$ is the minimum of the SMEFT potential. 
Replacing Eq. \ref{eq:SM-doublet} in $\mathcal{L}_{\mathrm{SMEFT}}$, one realizes that $h_{\text{\tiny S}}$ has non-canonical kinetic energy.
We thus use the field $h$, defined by:
\be
h \equiv h_s \bigg[1 + {v_T^4 C_{\mathcal{H}^6}^{(1)}\over 8 \, \Lambda^4}\bigg],
\ee
whose propagator is canonically normalized and whose mass we term $m_h$. An equivalent rescaling happens for $G^+_{\text{\tiny S}}$ and $G_{0,\text{\tiny S}}$.
The second aspect is that we need to write the relevant dependent parameters in terms of independent ones.%
\fn{These relations are in general different from those in the SM, due to terms proportional to powers of $1/\Lambda$.}
We take the following set of parameters as independent:
\be
G_F, M_W, M_Z, m_h, m_f,
\ee
besides all the WCs contained in Eqs. \ref{eq:Sall6} and \ref{eq:Sall8}.
The relevant dependent parameters are those occuring in our calculations, namely ${\overline {g}}_2, v_T, Y_f$ and $\lambda$. Now, ${\overline {g}}_2$ is the SU(2) gauge coupling occuring in the Lagrangian; we can determine it via the Fermi constant, $G_F$, which in turn can be obtained from muon decay; ignoring operators involving four fermion fields, we find:%
\fn{The relevant operators involving four fermion fields  are proportional to the Yukawa parameters of the muon and of the electron, so that their contribution can be safely neglected for our purposes.}
\be
{G_F\over \sqrt{2}} = { {\overline {g}}_2^2\over 8 M_W^2},
\ee
which can be inverted to yield:
\be
{\overline {g}}_2 = (\sqrt{2} G_F)^{1/2} \, 2 \, M_W.
\label{eq:g2}
\ee
To write $v_T$, one can see that the squared mass of the W boson in the Lagrangian of Eq. \ref{eq:SMEFT-Lag} is given by:
\be
M_W^2 = {{\overline{g}}_2^2 v_T^2\over 4} \biggl[ 1+{v_T^4 C_{{\mathcal{H}}^6}^{(1)}\over 4 \Lambda^4} \biggr],
\label{eq:MW2}
\ee
so that, combining this result with Eq. \ref{eq:g2}, we find:
\be
\label{eq:vT}
v_T = (\sqrt{2} G_F)^{-1/2} \bigg[1 - \dfrac{C_{{\mathcal{H}}^6}^{(1)}}{16 \, G_F^2 \, \Lambda^4}\bigg].
\ee
$Y_f$ can be determined via the fermion masses:
\be
m_f = \dfrac{v_T}{\sqrt{2}} \Big[ Y_f - \dfrac{ C_{f\mathcal{H}} v_T^2}{2 \, \Lambda^2} - \dfrac{C_{qf\mathcal{H}^5} v_T^4}{4 \Lambda^4} \Big];
\ee
combining this result with Eq. \ref{eq:vT}, we find:
\be
\label{eq:Yf}
Y_{f}=\sqrt{2} (\sqrt{2} G_{F})^{1/2} \, m_{f} + \frac{C_{f \mathcal{H}}}{2\sqrt{2}  \, G_{F} \Lambda^{2}} 
+\frac{C_{q f \mathcal{H}^{5}}}{8 \, G_{F}^{2} \Lambda^{4}}
+\frac{C_{ \mathcal{H}6}^{(1)} m_f}{4\sqrt{2} \, (\sqrt{2} G_{F})^{3/2} \Lambda^{4}}.
\ee
Finally, to determine $\lambda$, we minimize the potential to find the squared Higgs mass:
\be
m_h^{2} = 2 \lambda v_{T}^{2}\left(1-\frac{3 v_{T}^{2} C_{\mathcal{H}}}{2 \lambda \Lambda^{2}}-\frac{v_{T}^{4} C_{\mathcal{H}^6}^{(1)}}{4 \Lambda^{4}}-\frac{3 v_{T}^{4} C_{\mathcal{H}^8}}{2 \lambda \Lambda^{4}}\right),
\ee
so that, combining this result with Eq. \ref{eq:vT}, we find:
\be
\label{eq:lambda}
\lambda = \frac{G_{F} m_{h}^{2}}{\sqrt{2}} + \frac{3 C_{\mathcal{H}}}{2 \sqrt{2} G_{F} \Lambda^{2}} + \frac{3 C_{\mathcal{H}^{8}}}{4 G_{F}^{2} \Lambda^{4}} + \frac{C_{\mathcal{H}^{6}}^{(1)} m_{h}^{2}}{4 \sqrt{2} G_{F} \Lambda^{4}}.
\ee
Using this relation, we can also conclude that the coefficient of the tri-linear Higgs coupling, 
 which we term $\lambda_3$ (we have $V \ni {m_h^2\over 2} h^2+\lambda_3 h^3+\lambda_4 h^4$), is given by:
\be
\label{eq:lambda3}
\lambda_3 = \frac{m_h^2}{2} (\sqrt{2} G_{F})^{1/2} - \frac{C_{\mathcal{H}} (\sqrt{2} G_{F})^{-3/2}}{\Lambda^{2}} - \frac{2 C_{\mathcal{H}^8} (\sqrt{2} G_{F})^{-5/2}}{\Lambda^{4}},
\ee
which shall also be useful below.

\section{Fit procedure and  SMEFT results}
\label{sec:fit-pro}

In this section, we describe our fit procedure. We perform $\chi^2$ fits to the global Higgs signal strengths --- not only in the context of SMEFT, but also in that of the 2HDM. This allows us to compare each of these two approaches with the experimental results, as well as with each other; the last aspect thus enables us to ascertain the quality with which the EFT presented in the previous section can describe the 2HDM introduced in Section \ref{sec:2hdm}. In what follows, we specify the fermion flavors whenever it is relevant.

\subsection{Direct determination of Higgs couplings}

\begin{figure}[h!]
\centering
\subfloat{\includegraphics[width=0.25\linewidth]{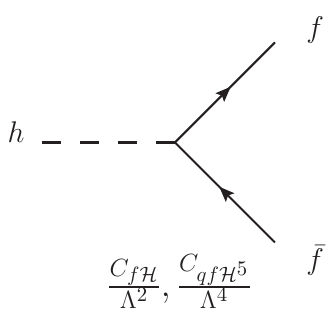}}
\quad
\subfloat{\includegraphics[width=0.3\linewidth]{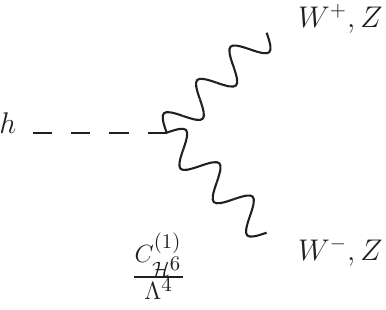}}
\subfloat{\includegraphics[width=0.25\linewidth]{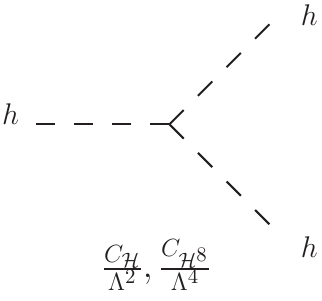}}
\quad
\caption{Feynman diagrams and the coefficients that contribute to tree level Higgs decays at 
${\cal{O}}({1\over \Lambda^2})$ and  ${\cal{O}}({1\over \Lambda^4})$.}
\label{fig:feyns}
\end{figure}

In Fig. \ref{fig:feyns}, we schematically show which coefficients contribute to the tree level Higgs couplings
at ${\cal{O}}({1\over \Lambda^2})$ and  ${\cal{O}}({1\over \Lambda^4})$.  It is apparent that, at dimension-6,
there is no contribution from the Higgs decays to vector bosons.  We perform our fit to Higgs decays
including only the coefficients generated in the 2HDM.    In practice, it could be that a global fit to data (Higgs, top, electroweak precision, and di-boson production) measures some pattern of non-zero coefficients, which might suggest an underlying 2HDM model.  In this scenario, a subsequent fit could be done just to the coefficients of the 2HDM.

The Higgs signal strengths may generically be represented  as $\mu^P_{pp \to h \to f}$, where $P$ represents a production mode and $f$ a final state. Using the narrow width approximation, we write:
\be
\mu^P_{pp \to h \to f}
=
\mu_P \times \mu_f
=
{\sigma^P(pp \rightarrow h)\over \sigma^P(pp \rightarrow h)_{\mathrm{SM}}}
\times
{\mathrm{BR}(h\rightarrow f)\over \mathrm{BR}(h \rightarrow f)_{\mathrm{SM}}} .
\label{eq:signalstrength}
\ee
We then compare the different signal strengths evaluated theoretically (both for the SMEFT and the 2HDM) with those measured experimentally. We consider the  $ggh$, VBF, $Wh$, $Zh$ and $t{\overline{t}}h$  production modes, and  the $\gamma \gamma$, $b{\overline{b}}$, $\tau^+\tau^-$, $W^+W^-$ and $ZZ$  final states. For the SM production modes, we use the masses and couplings from Ref. \cite{ParticleDataGroup:2020ssz} (with $m_h=125$ GeV); for the SM branching ratios (BRs), we use the  results computed by the LHC Higgs cross section working group \cite{https://doi.org/10.23731/cyrm-2017-002}.
For the experimental signal strengths, we include the combined 7/8 TeV ATLAS/CMS data \cite{ATLAS:2016neq}, the $13$ TeV and $139 ~\mathrm{fb}^{-1}$ data from ATLAS \cite{ATLAS:2020qdt}, and the $13$ TeV and $137 ~\mathrm{fb}^{-1}$ data from CMS \cite{CMS:2020gsy} (including known correlations). 

\begin{figure}[h!]
\centering
\subfloat{\includegraphics[width=0.49\linewidth]{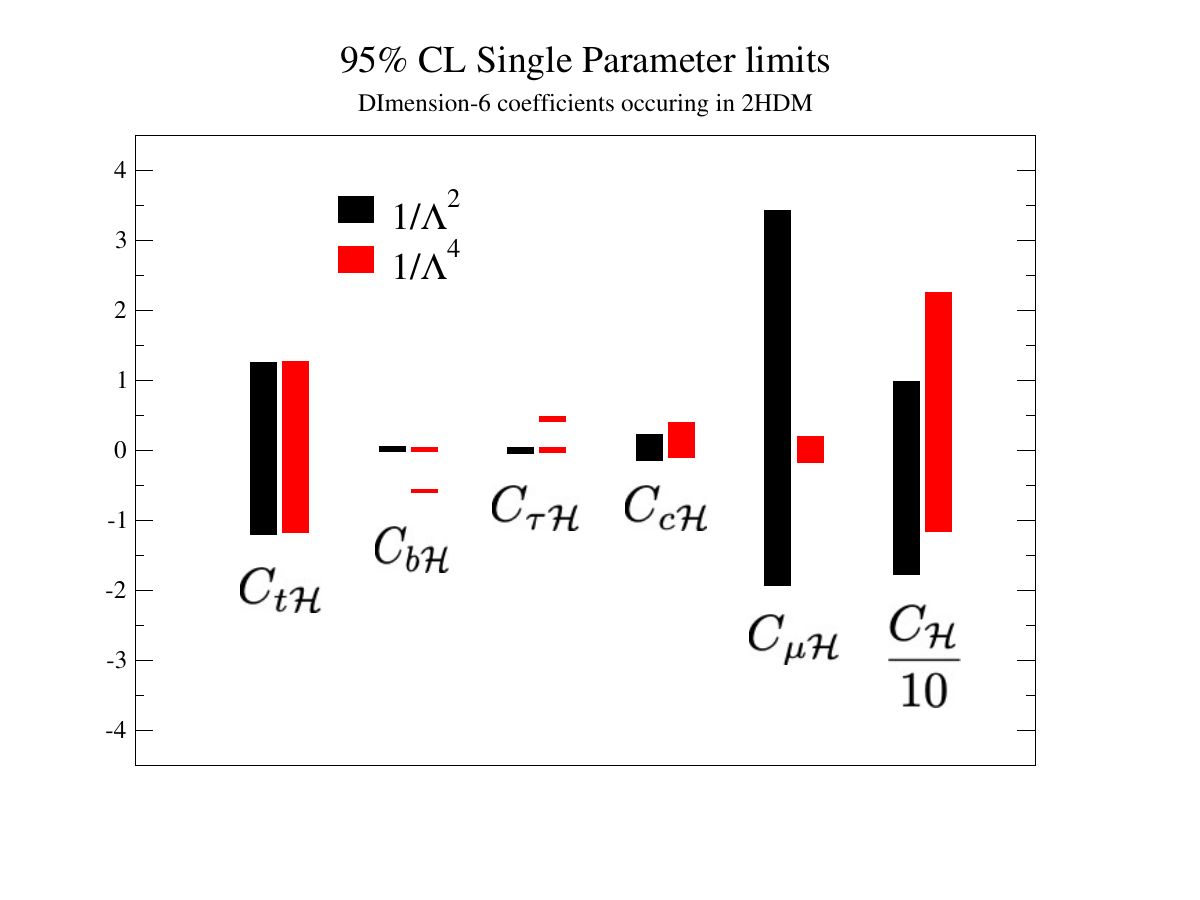}}
\subfloat{\includegraphics[width=0.49\linewidth]{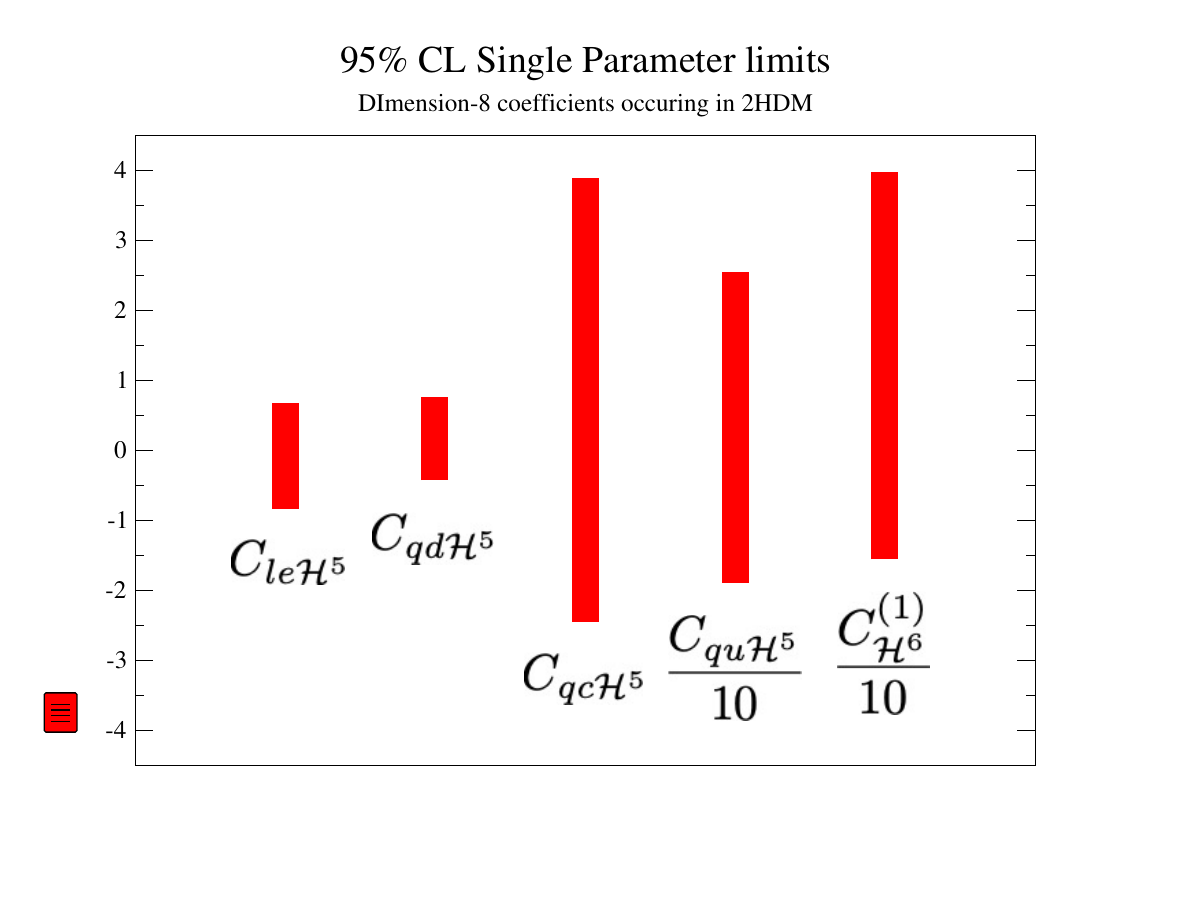}}
\caption{Single parameter limits on WCs occuring in the 2HDM at 
${\cal{O}}({1\over\Lambda^2})$ and ${\cal{O}}({1\over\Lambda^4})$ with $\Lambda=1$~TeV.}
\label{fig:Bar}
\end{figure}

\begin{figure}[h!]
\centering
\subfloat{\includegraphics[width=0.55\linewidth]{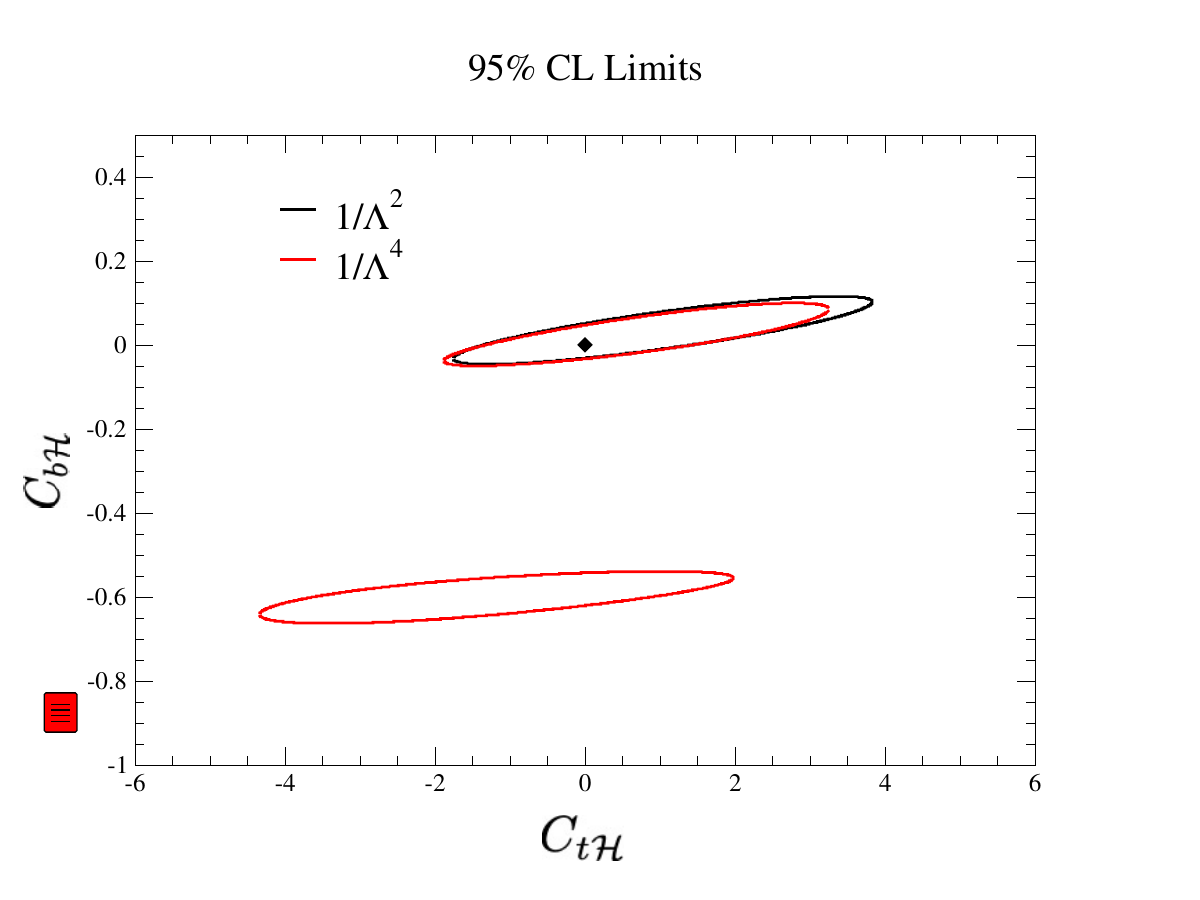}}
\caption{{Two parameter fit to $C_{t\mathcal{H}}-C_{b\mathcal{H}}$ from single Higgs production with $\Lambda=1~$TeV}. The diamond represents the SM value.}
\label{fig:3-Corr}
\end{figure}

We begin by computing the $95\%$ confidence level (CL) limits on the WCs that are generated in the 2HDM. 
Fig. \ref{fig:Bar} compares the limits when Eq. \ref{eq:signalstrength} is expanded consistently to ${\cal{O}}({1\over\Lambda^2})$ with 
those resulting from an expansion to ${\cal{O}}({1\over\Lambda^4})$.  We note that for poorly constrained coefficients, for example $C_{\mathcal{H}}$ and $C_{\mu{\mathcal{H}}}$,  the limits are quite sensitive to the expansion.
Moreover, in the case of $C_{b{\mathcal{H}}}$ and $C_{\tau{\mathcal{H}}}$, there are two separate solutions when ${\cal{O}}({1\over\Lambda^4})$ effects are included, but only one when these effects are excluded; the extra solution shall correspond to the so-called \textit{wrong-sign solution} in the context of the 2HDM, as shall be discussed below.
In Fig. \ref{fig:3-Corr}, we consider the limits in the $C_{t\mathcal{H}}-C_{b\mathcal{H}}$ plane, 
with other coefficients set to zero.
As before, we find two disconnected solutions for $C_{b{\mathcal{H}}}$ only when ${\cal{O}}({1\over\Lambda^4})$ effects are included. In both solutions, there is a large correlation between $C_{t\mathcal{H}}$ and $C_{b\mathcal{H}}$. It is also clear that $C_{t\mathcal{H}}$ is much less contrained in any of the solutions than it was in Fig. \ref{fig:Bar}.

\subsection{Indirect determination of Higgs tri-linear coupling}
\label{sec:tri-linear}

Finally, we are interested in ascertaining the importance of the Higgs self-interactions. At leading order, direct experimental constraints on those interactions can only be obtained from double Higgs production; the cross section for this type of process, however, is much smaller than that for single Higgs production. The authors of Ref. \cite{Degrassi:2016wml,Degrassi:2021uik } thus propose to indirectly determine the Higgs self-interactions by calculating the contribution of such interactions using higher order corrections to single Higgs production and decays. These contributions depend on the cubic self-interaction, which in our SMEFT framework is given by Eq. \ref{eq:lambda3} and depends on the WCs $C_{\mathcal{H}}$ and $C_{\mathcal{H}^8}$. For each signal strength $\mu_{j}$, we adopt the prescription of Ref. \cite{Degrassi:2016wml} to calculate the effects of the Higgs tri-linear self-interactions on $\mu_{j}$; we term those effects $\delta \hat{\mu}_j$.
The combination of $\mu_{j}$ and  $\delta \hat{\mu}_j$ is ambiguous at ${\cal{O}}({1\over\Lambda^4})$, since we do not know the interference effects between the two types of contributions. We combine them as
\be
\mu_j^{\mathrm{com}} = \mu_j + \delta {\hat{\mu_j}}.
\ee
It should be clear that this is only a rough estimation of the combined contribution of the Higgs signal strenth of Eq.  \ref{eq:signalstrength} and the Higgs self-interactions.

\section{Results}
\label{sec:res}

We now turn to our main results. We start by discussing in section \ref{sec:Expansions} two relevant aspects related to the validity of the EFT expansion; after that, we present our fits in section \ref{sec:Extraction}. In both these sections, we ignore the effects on the signal strengths coming from the Higgs self-interactions; these shall be investigated in section \ref{sec:tri}. In what follows, and wherever the results explicitly depend on the scale $\Lambda$, we take by default $\Lambda = 1$ TeV.

\subsection{Preliminary aspects} 
\label{sec:Expansions}

\begin{figure}[h!]
\centering
\includegraphics[width=0.97\linewidth]{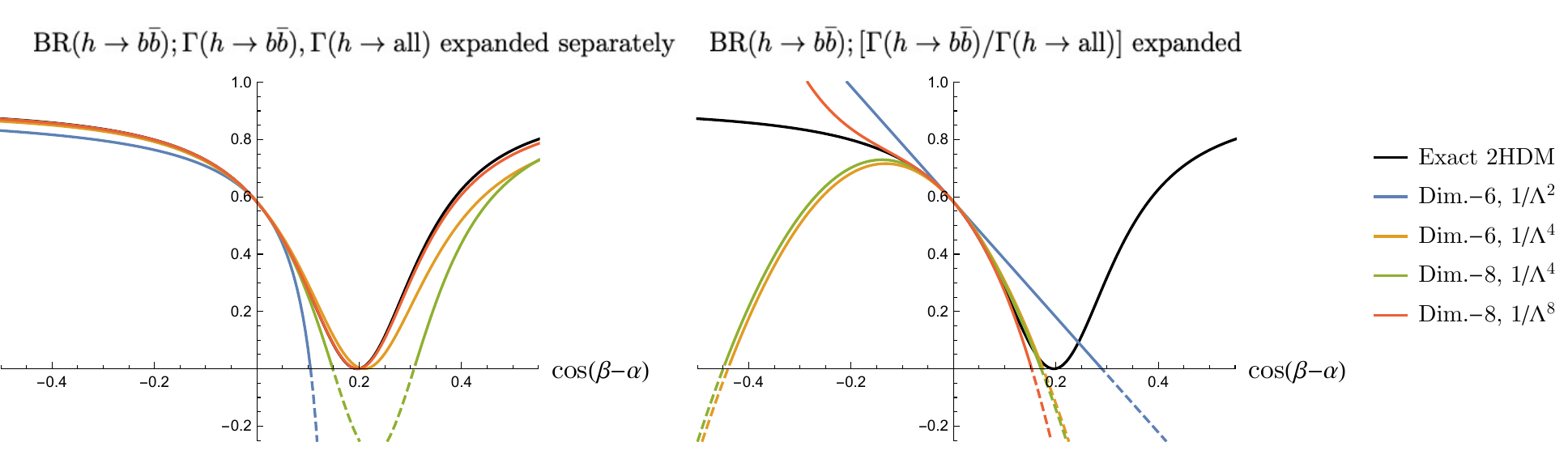}
\caption{Calculations of $\mathrm{BR}(h\rightarrow bb)$ in the Type-II 2HDM for $\tan\beta=5$ with two different expansions: on the left, the partial width of $h \rightarrow b\bar{b}$ and the total width of $h$ are expanded separately to the given order in the EFT; on the right, the ratio of the partial width to the total width is expanded.}
\label{fig:hbb-BR}
\end{figure}

The first aspect we discuss here concerns the existence of different methods of expanding a BR.
Let us consider the BR of $h \rightarrow b\bar{b}$ in the Type-II 2HDM, given by $
\mathrm{BR}(h \rightarrow b\bar{b}) = \Gamma({h \rightarrow b\bar{b}}) / \Gamma({h \rightarrow \mathrm{all}})$. When expanding the EFT results in powers of $\frac{1}{\Lambda^2}$, one may choose either to expand separately $\Gamma({h \rightarrow b\bar{b}})$ and $\Gamma({h \rightarrow \mathrm{all}})$, or to expand $\mathrm{BR}(h \rightarrow b\bar{b})$ as a whole. The crucial difference between the two approaches is that the former expands $\Gamma({h \rightarrow \mathrm{all}})$ itself, whereas the latter expands the \textit{inverse} of $\Gamma({h \rightarrow \mathrm{all}})$. Now, when the contributions from higher order operators to $\Gamma({h \rightarrow \mathrm{all}})$ are sizable compared to the SM contribution, the expansion of $1/\Gamma({h \rightarrow \mathrm{all}})$ does not converge.%
\fn{This is not a statement about a lack of convergence of the EFT; rather, it is a general statement about the radius of convergence of the power series of rational functions.}
As a consequence, the expansion of $\mathrm{BR}(h \rightarrow b\bar{b})$ as a whole also does not converge. This is what can be seen on the right panel of Fig.~\ref{fig:hbb-BR}, for almost the entire range of values of $\cos({\beta - \alpha})$. In order to properly investigate these regions, then, one should expand $\Gamma({h \rightarrow \mathrm{all}})$ itself and leave it in the denominator, as done on the left panel. 

Still concerning Fig. \ref{fig:hbb-BR}, it is clear in both plots that there are truncations which lead to BRs larger than 1 or smaller than 0
(which is clearly unphysical, signalling that the expansion is not valid).
In particular, the green curve on the left plot --- that includes linear effects from EFD 8 operators, but neglects the squared effects -- describes negative BRs for a certain range of values of $\cos(\beta-\alpha)$.%
\fn{This is a general behaviour in SMEFT truncations whenever there is a removal of squared effects.}
On the other hand, and as shall be seen below, such range of values (in Type-II) is excluded by experiment, as $|\cos({\beta - \alpha})|$ is required to be very close to zero. Therefore, in what follows, we consistently ignore squared effects from EFD 8 operators, which are of order ${\cal{O}}({1\over \Lambda^8})$.

\begin{figure}[h!]
\centering
\includegraphics[width=0.55\linewidth]{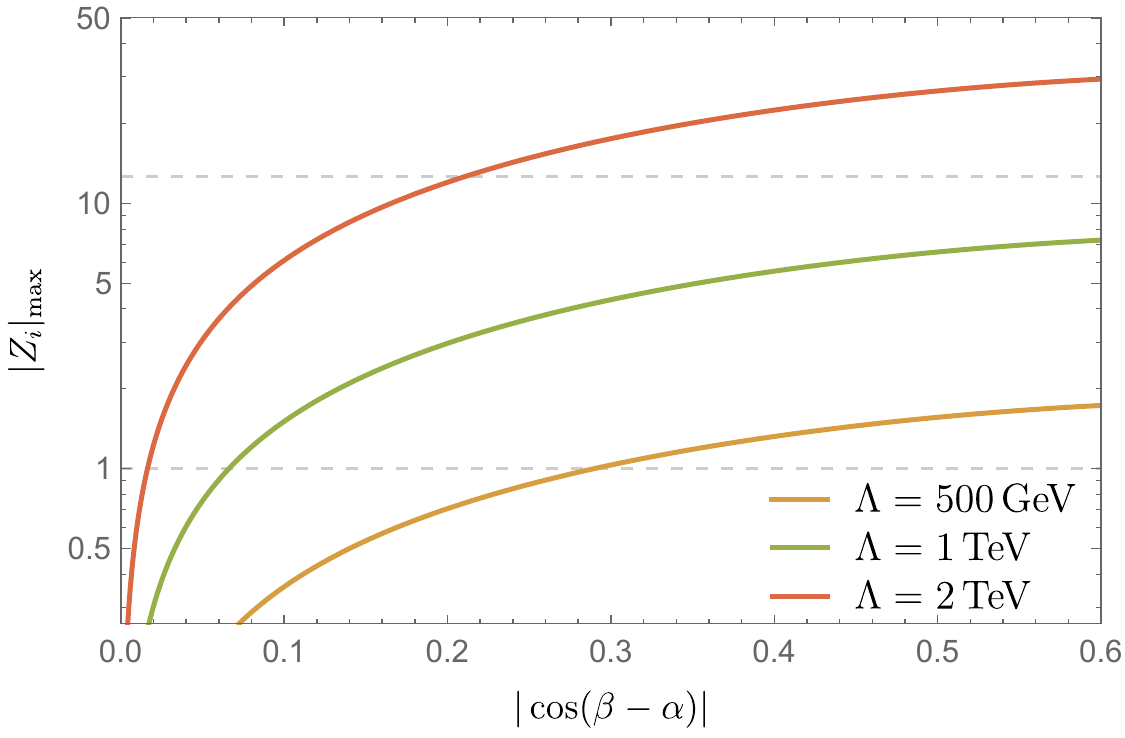}
\caption{The maximum absolute value among the different $Z_i$ against $\cos(\beta-\alpha)$, for three different scales. The dashed lines highlight the values 1 and $4 \pi$.}
\label{fig:unit}
\end{figure}

The second preliminary aspect is related to the range of validity for the EFT expansion of the 2HDM. As discussed in section \ref{sec:decoupling}, the EFT requires the decoupling limit, which in turn requires the different $Z_i$ parameters of eq. \ref{eq:theZs-real} to obey $Z_i/(4 \pi) \simeq \mathcal{O}(1)$. Let us then define $|Z_i|_{\mathrm{max}}$ as the maximum absolute value among the different $Z_i$, for a certain point in the parameter space. In Fig. \ref{fig:unit}, we show the $|Z_i|_{\mathrm{max}}$ as a function of $|\cos(\beta-\alpha)|$, for three different values of the scale $\Lambda$.
It is clear that the green and orange curves are always below the upper dashed line, which corresponds to $|Z_i|_{\mathrm{max}} = 4 \pi$. In other words, for the two lower scales (500 GeV and 1 TeV), the entire range of $\cos(\beta-\alpha)$ shown provides a valid description of the EFT expansion with regards to perturbative unitarity, since it is compatible with the requirement $Z_i/(4 \pi) \simeq \mathcal{O}(1)$. The heavier scale $\Lambda = 2$ TeV starts yielding $Z_i/(4 \pi) > 1$ for $\cos(\beta-\alpha) \gtrsim 0.2$.

\subsection{Extraction of Higgs couplings and matching to 2HDM} 
\label{sec:Extraction}
\begin{figure}[h!]
\centering
\subfloat{\includegraphics[width=0.45\linewidth]{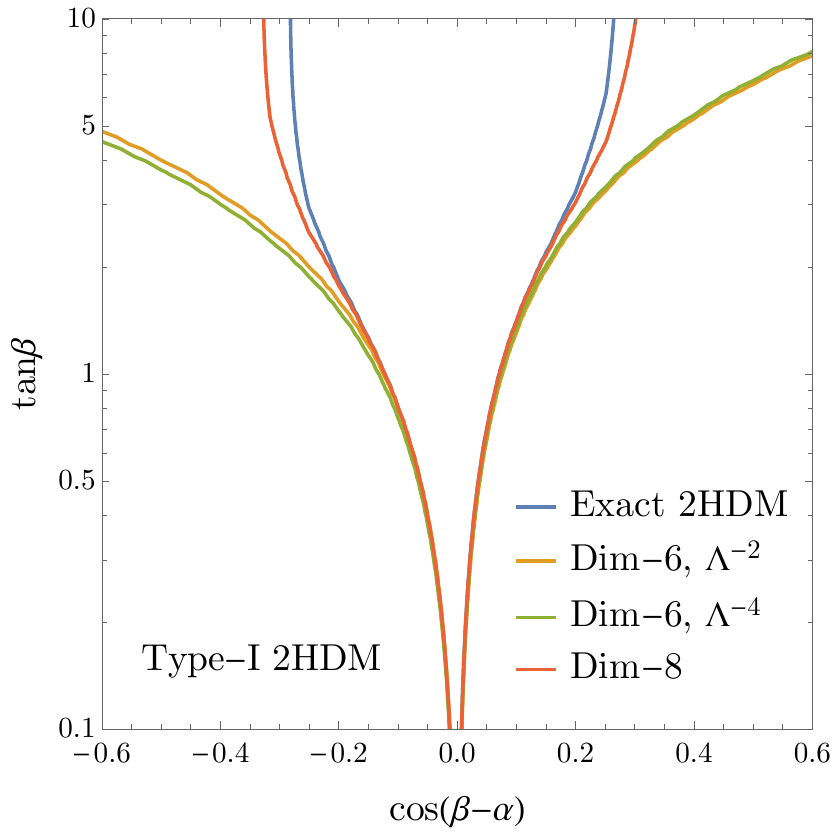}}
\quad
\subfloat{\includegraphics[width=0.45\linewidth]{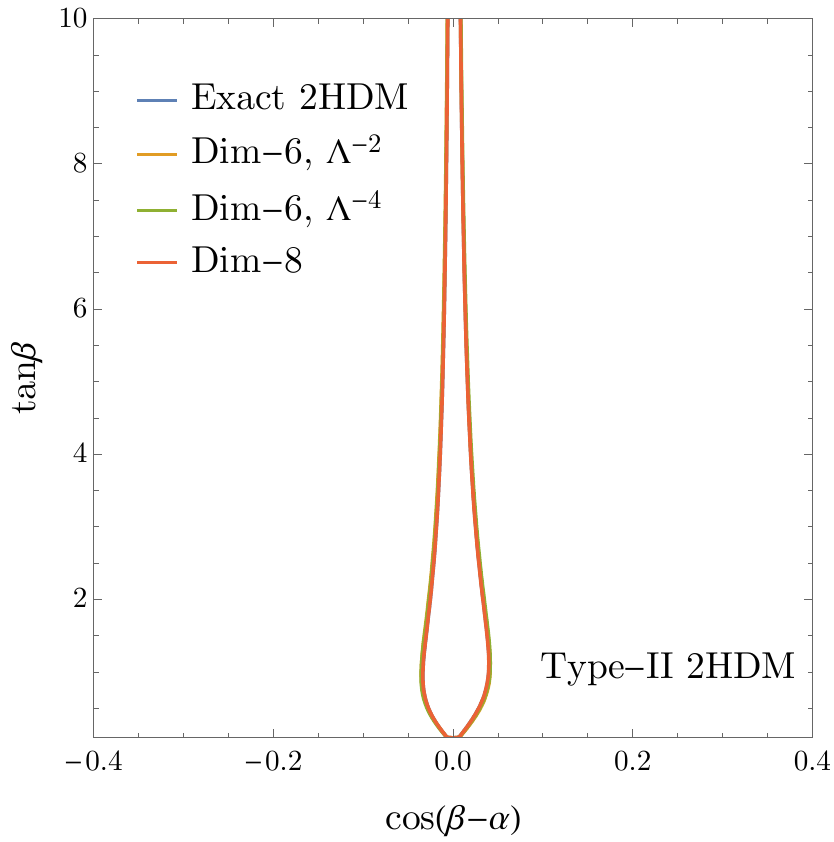}}\\
\subfloat{\includegraphics[width=0.45\linewidth]{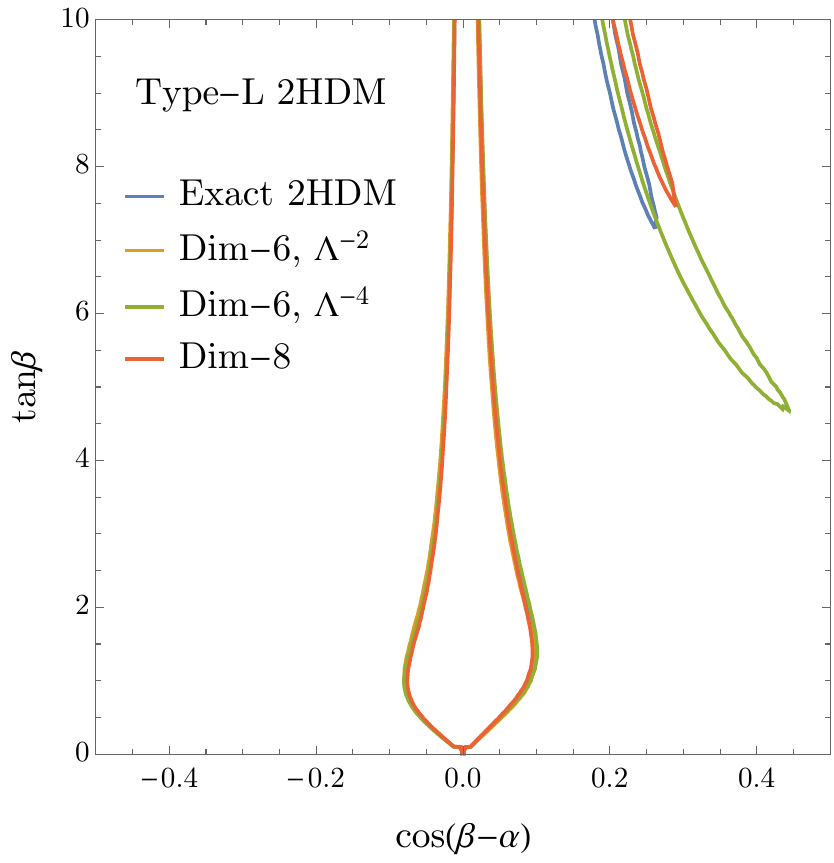}}
\quad
\subfloat{\includegraphics[width=0.45\linewidth]{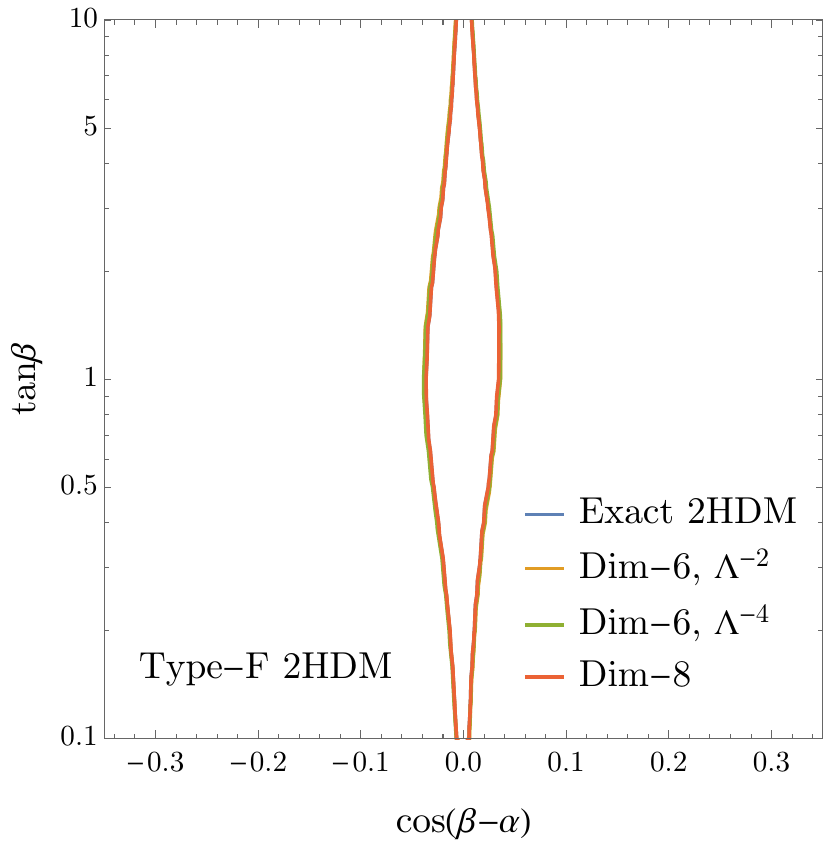}}
\caption{95$\%$ CL constraints from the LHC data on the $\tan \beta$ vs $\cos({\beta - \alpha})$ plane, for Type-I (top left), Type-II (top right), Type-L (bottom left) and Type-F (bottom right). The blue lines represent limits on the full UV model, whereas the remaining lines describe limits within the SMEFT framework in different approximations made at the level of the Higgs signal strengths.}
\label{fig:4-types}
\end{figure}
In Fig. \ref{fig:4-types}, we present the fits to both the exact 2HDM and the SMEFT matched to the 2HDM, for  the four types of 2HDM. In order to ascertain the importance of higher EFD operators in the EFT description, 
we consider three different approximations for the calculation of the Higgs signal strengths in the SMEFT: with EFD 6 operators and excluding squared contributions (orange), with EFD 6 operators but including squared contributions (green), and with EFD 8 operators (red).
In what follows, we discuss in detail each type of 2HDM.

Let us begin with Type-I. Here, the results for the SMEFT up to EFD 6 are weakly constrained, especially for higher values of $\tan \beta$. As already noted in Ref. \cite{Belusca-Maito:2016dqe}, this is due to a combination of two reasons: first, up to EFD 6 (and excluding self-interactions), the only WCs contributing to the SMEFT fit are those modifying the Yukawa couplings, namely $C_{f\mathcal{H}}$ (recall Eq. \ref{eq:Sall6}); second, the Yukawa parameters $\eta_f$ in Type-I are equal to 1 for all types of fermions (see Table \ref{tab:types}), which implies that the $C_{f\mathcal{H}}$ up to EFD 6 (e.g. in the first term of the right-hand side of Eq. \ref{eq:CuH}) are suppressed by $\tan\beta$. As a consequence, for high values of $\tan \beta$, the EFT for the 2HDM Type-I truncated with EFD 6 operators has no relevant information that can be restricted by experiment. This is in clear contrast with the full Type-I 2HDM, which obviously contains more predictions than simply the Yukawa interactions (in particular, it contains those related to the interactions between the Higgs and gauge bosons), and thus ends up being contrained by the experimental results. Therefore, the region of parameter space of the Type-I 2HDM where $\tan \beta$ takes high values constitutes a scenario for which the EFT truncated with EFD 6 operators is utterly unable to provide a correct description of the full UV model. The inclusion of quadratic effects does not change this picture, as it does not alter the two reasons given above. 

What \textit{does} change the picture --- and quite significantly --- is the inclusion of EFD 8 operators. Indeed, the EFT now has information besides the Yukawa couplings; in particular, it contains a prediction for the cubic Higgs-gauge interactions (via the coefficient $C_{\mathcal{H}^6}^{(1)}$, cf. Eq. \ref{eq:Sall8}), which is not supressed with $\tan \beta$ (see Eq. \ref{eq:46b}). Accordingly, the SMEFT framework is now able to be experimentally constrained for the entire spectrum of $\tan \beta$. Not surprisingly, then, it now provides a very accurate description of the full UV model, as can be seen in the figure. Finally, we note that, although there is an explicit scale dependence when the EFD 8 operators are included, the dependence is not numerically significant. 

Concerning the other types of 2HDM, the figure shows that, for Type-II and Type-F, the SMEFT approach provides an excellent description of the corresponding full UV model, even if one truncates the expansion with EFD 6 operators and neglects quadratic effects in the signal strengths. The difference between these models and Type-I is that, in the former, there is always at least one fermion type whose $\eta_f$ parameter is proportional to $\tan^2 \beta$, so that the respective $C_{f\mathcal{H}}$ no longer scales with $\tan^{-1} \beta$ --- which in turn implies that the SMEFT is constrained by experiment for high values of $\tan \beta$, even if it only contains the Yukawa-related WCs. For these two types of 2HDM, therefore, the inclusion of higher-order effects in the SMEFT expansion is irrelevant.

Finally, Type-L stands out among the other types, as it is still compatible with the interesting scenario of the wrong-sign solution, corresponding to the isolated band centered around $\cos({\beta-\alpha})=0.3$.%
\fn{Although Type-II and Type-L would in principle allow such a solution, it is ruled out by the most recent experimental data\cite{Atkinson:2021eox}.}
Here, we verify what was already observed in Ref. \cite{Belusca-Maito:2016dqe}, namely: the wrong-sign solution cannot be captured by the SMEFT description if only the linear effects of the EFD 6 operators are included in the SMEFT predictions for the Higgs signal strengths. This is because the likelihood that results from such an approximation is a Gaussian one, which contains only one minimum; this minimum corresponds to the solution where the WCs are close to zero, which is the solution favored by the experimental data. So, in order to obtain different minima, the likelihood must be non-Gaussian, which in turn can only be obtained by including the higher order quadratic effects.
Accordingly, the wrong-sign solution in Type-L constitutes a scenario where a SMEFT approach that consistently includes only the $1/\Lambda^2$ terms in the Higgs signal strengths completely misses the description of the full model. In order to capture that region, though, one does not necessarily need to include the effects of EFD 8 operators: the figure shows that the squared terms of EFD 6 operators already leads to the generation of the wrong-sign solution.
On the other hand, it is also clear from the figure that such a solution is far from being a faithful reproduction of the band of the full UV model; in fact, whereas the latter only reaches values of $\cos({\beta-\alpha})$ slightly larger than $0.2$, the green  band extends to values larger than $0.4$. The reason is that, in the full UV model, the large values of $\cos({\beta-\alpha})$ are ruled out by measurements of the Higgs couplings to gauge bosons; but since the green band does not have information about such couplings (they only show up with $C_{\mathcal{H}^6}^{(1)}$, at higher order), it is not constrained to small values of $\cos({\beta-\alpha})$. This simple reasoning is confirmed by the orange band, which shows that the SMEFT result is already restricted to smaller values of  $\cos({\beta-\alpha})$ when the SMEFT expansion is truncated with EFD 8 operators, thus reproducing quite well the solution of the full model.

As a final note, we should stress that the wrong-sign solution does not spoil the convergence of the EFT expansion \cite{Belusca-Maito:2016dqe}. This is not obvious, since that solution requires the EFT effects to be twice as large (in modulus) as those of the SM. On the other hand, the validity of the EFT expansion in Eq. \ref{eq:basics} does not necessarily require dimension-6 operators to have smaller effects than those with dimension 4, but only that the subsequent orders do not become relevant. Actually, there are many examples where dimension-6 operators can be much larger than dimension-4 operators, without ruining the EFT expansion \cite{Contino:2016jqw}. As we discussed above, the wrong-sign solution described by SMEFT in Type-L reproduces quite well the full model, which implies the convergence of the EFT expansion.

\subsection{Inclusion of Higgs tri-linear couplings}
\label{sec:tri}

\begin{figure}[h!]
\centering
\subfloat{\includegraphics[width=0.45\linewidth]{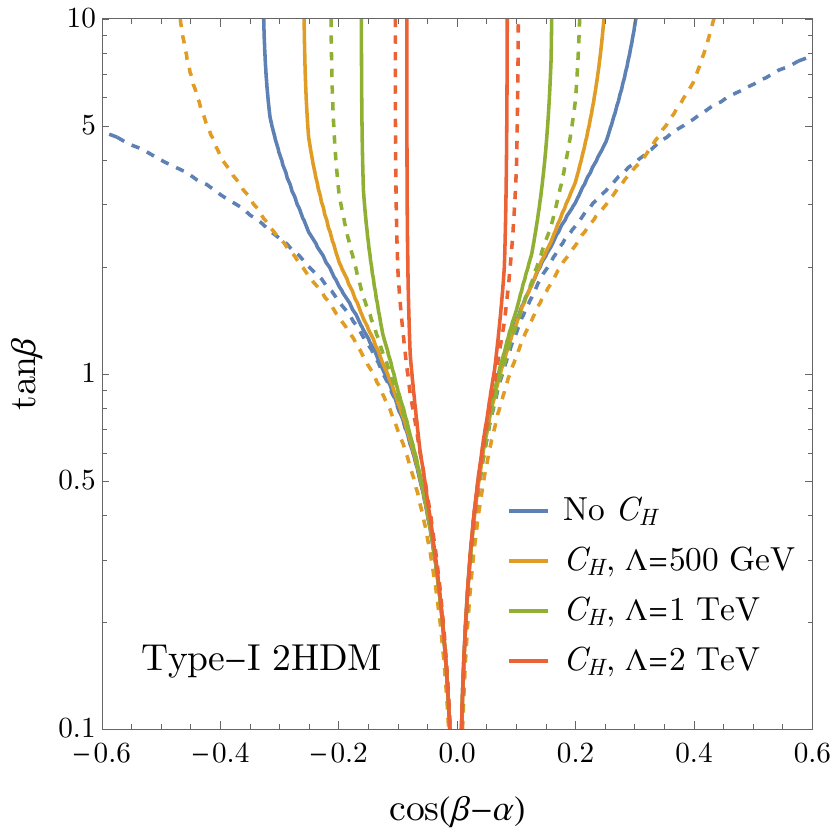}}
\quad
\subfloat{\includegraphics[width=0.45\linewidth]{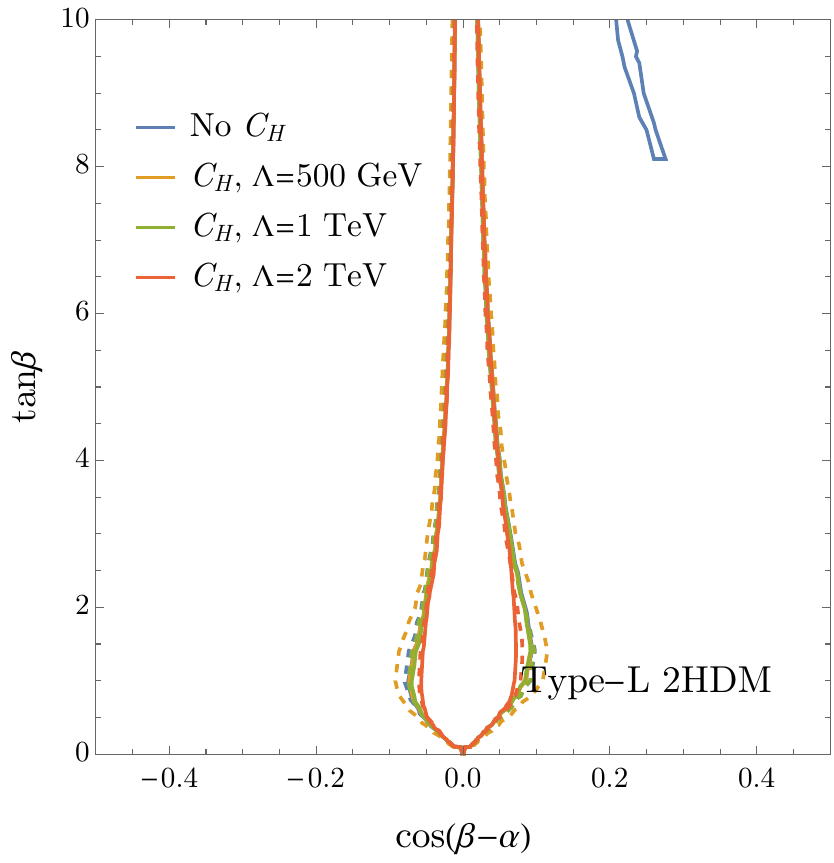}}
\caption{95$\%$ CL constraints from the LHC data on the $\tan \beta$ vs $\cos({\beta - \alpha})$ plane, for Type-I (left) and Type-L (right) and including the effects of the Higgs tri-linear coupling. The different types of lines represent different approximations through which the Higgs signal strengths are calculated: by keeping only linear effects of the EFD 6 operators (dashed lines) or by including up to EFD 8 operators (solid lines).}
\label{fig:2-types-with-CH}
\end{figure}

We now investigate the effects of including the Higgs self-interactions in the SMEFT predictions for the Higgs signal strengths, as described in section \ref{sec:tri-linear}. In Fig. \ref{fig:2-types-with-CH}, we show again the fits for Type-I and Type-L, but now explicitly comparing the results with and without the Higgs self-interactions. The differences are striking and motivate a complete calculation which would include the neglected interference effects in the SMEFT predictions. 
In Type-I up to EFD 6, the curves that include the self-interactions no longer have the problem of describing a prediction that is globally suppressed with $\tan \beta$; precisely due to the self-interactions, indeed, such a prediction also includes $C_{\mathcal{H}}$, which does not depend on $\tan \beta$ (cf. Eq. \ref{eq:cH}). It does depend, however, very strongly on the scale $\Lambda$,  as can also be seen in the figure. 
In the case of Type-L, the inclusion of the self-interaction excludes the wrong-sign solution, and there is no relevant dependence on $\Lambda$.
We do not compare these results with the full 2HDM, because this would require comparing
with the one-loop predictions for the 2HDM in the heavy mass limit, which is beyond the scope
of this work.  Fig. \ref{fig:2-types-with-CH} motivates such a comparison.
Finally, for the entire range of $\cos(\beta-\alpha)$ considered, all the results in Fig. \ref{fig:2-types-with-CH} respect the requeriment of perturbative unitarity. In fact, and as can be seen comparing to Fig. \ref{fig:unit}, never does one scale reach the region of values of $\cos(\beta-\alpha)$ where $Z_i/(4 \pi) > 1$.

\section{Conclusions}
\label{sec:conc}

This work considers  the SMEFT as an effective low-energy theory for the 2HDM. Whereas the usual approach to describe a full UV model with the SMEFT truncates the EFT expansion of the Lagrangian with $\mathcal{O}({1\over \Lambda^2})$ terms, we  truncate including the dimension-8 $\mathcal{O}({1\over \Lambda^4})$ terms and investigate the importance of these terms. Assuming the decoupling limit, we integrate out the heavy scalar doublet $H_2$ of the 2HDM, which allows us to consistently derive the matching between the SMEFT and the 2HDM up to $\mathcal{O}({1\over \Lambda^4})$. This matching is used to convert the LHC constraints on the SMEFT Wilson coefficientss into constraints on the parameter space of the 2HDM, and thus evaluate the quality with which the SMEFT is able to reproduce the LHC direct constraints on the 2HDM and the importance of the $\mathcal{O}({1\over \Lambda^4})$ terms.

Focusing on the scenario where CP is conserved in the scalar sector, we analyze the four 
types of 2HDM models that result from different applications of the $Z_2$ symmetry to the fermion sector, using the latest experimental data to perform fits for the UV complete version.
 We find two situations where the inclusion of the  $\mathcal{O}({1\over \Lambda^4})$ terms is crucial: the region in Type-I with moderate values of $\tan \beta$, and the wrong-sign solution found in Type-L. In both cases, the SMEFT approach provides a very poor description if truncated at $\mathcal{O}({1\over \Lambda^2})$, but a very good one if the $\mathcal{O}({1\over \Lambda^4})$ terms are included. The reason for this behaviour is that the Higgs-gauge interactions that constrain the full 2HDM  only show up in the SMEFT expansion with the $\mathcal{O}({1\over \Lambda^4})$ terms. For the remaining 2HDM models, the SMEFT truncated at $\mathcal{O}({1\over \Lambda^2})$  provides a good description of the UV complete model, so that the inclusion of $\mathcal{O}({1\over \Lambda^4})$ terms becomes irrelevant.

We also discuss the effects of the self-Higgs interactions that arise beyond tree level  in  single Higgs production and decay in the SMEFT. These lead to strong constraints on both Type-I and Type-L models in the SMEFT,  along with a numerically significant scale dependence. It would be of interest to compute 
the SMEFT predictions for the signal strengths  to one-loop order (two loop for gluon fusion) to assess the relevance of the Higgs-self couplings in a consistent manner including the interference between the contributions.  The loop suppression factor of $1/(16 \,  \pi^2)$ could have similar size effects as the dimension-8 contributions which scale as ${v^4\over \Lambda^4}$.
Also of interest would be to compare our results with those of a different approach, namely the Higgs Effective Field Theory (HEFT); there, instead of the doublet $H_2$, one integrates out the heavy mass states of the 2HDM. The validity of HEFT is expected to be more general than that of SMEFT.

\section*{Acknowledgements}
We thank Adam Falkowski, Xiaochuan Lu, Christopher Murphy and Robert Szafron for discussions.
SD, DF,  and MS are supported by the United States Department of Energy under Grant Contract DE- SC0012704. 
SH is supported in part by the DOE Grant DE-SC0013607, and in part by the Alfred P. Sloan Foundation Grant No. G-2019-12504. Digital data is posted at \url{https://quark.phy.bnl.gov/Digital_Data_Archive/dawson/dim8_22}.

\bibliographystyle{utphys}
\bibliography{MyReferences.bib}

\providecommand{\href}[2]{#2}\begingroup\raggedright\begin{thebibliography}{10}

\bibitem{Brivio:2017vri}
I.~Brivio and M.~Trott, ``{The Standard Model as an Effective Field Theory},''
  \href{http://dx.doi.org/10.1016/j.physrep.2018.11.002}{{\em Phys. Rept.}
  {\bfseries 793} (2019) 1--98},
  \href{http://arxiv.org/abs/1706.08945}{{\ttfamily arXiv:1706.08945
  [hep-ph]}}.

\bibitem{Almeida:2021asy}
E.~d.~S. Almeida, A.~Alves, O.~J.~P. \'Eboli, and M.~C. Gonzalez-Garcia,
  ``{Electroweak legacy of the LHC run II},''
  \href{http://dx.doi.org/10.1103/PhysRevD.105.013006}{{\em Phys. Rev. D}
  {\bfseries 105} no.~1, (2022) 013006},
  \href{http://arxiv.org/abs/2108.04828}{{\ttfamily arXiv:2108.04828
  [hep-ph]}}.

\bibitem{Ethier:2021bye}
{\bfseries SMEFiT} Collaboration, J.~J. Ethier, G.~Magni, F.~Maltoni,
  L.~Mantani, E.~R. Nocera, J.~Rojo, E.~Slade, E.~Vryonidou, and C.~Zhang,
  ``{Combined SMEFT interpretation of Higgs, diboson, and top quark data from
  the LHC},'' \href{http://dx.doi.org/10.1007/JHEP11(2021)089}{{\em JHEP}
  {\bfseries 11} (2021) 089}, \href{http://arxiv.org/abs/2105.00006}{{\ttfamily
  arXiv:2105.00006 [hep-ph]}}.

\bibitem{deBlas:2021wap}
J.~de~Blas, M.~Ciuchini, E.~Franco, A.~Goncalves, S.~Mishima, M.~Pierini,
  L.~Reina, and L.~Silvestrini, ``{Global analysis of electroweak data in the
  Standard Model},'' \href{http://arxiv.org/abs/2112.07274}{{\ttfamily
  arXiv:2112.07274 [hep-ph]}}.

\bibitem{Ellis:2020unq}
J.~Ellis, M.~Madigan, K.~Mimasu, V.~Sanz, and T.~You, ``{Top, Higgs, Diboson
  and Electroweak Fit to the Standard Model Effective Field Theory},''
  \href{http://dx.doi.org/10.1007/JHEP04(2021)279}{{\em JHEP} {\bfseries 04}
  (2021) 279}, \href{http://arxiv.org/abs/2012.02779}{{\ttfamily
  arXiv:2012.02779 [hep-ph]}}.

\bibitem{Alasfar:2020mne}
L.~Alasfar, A.~Azatov, J.~de~Blas, A.~Paul, and M.~Valli, ``{$B$ anomalies
  under the lens of electroweak precision},''
  \href{http://dx.doi.org/10.1007/JHEP12(2020)016}{{\em JHEP} {\bfseries 12}
  (2020) 016}, \href{http://arxiv.org/abs/2007.04400}{{\ttfamily
  arXiv:2007.04400 [hep-ph]}}.

\bibitem{DeBlas:2019qco}
J.~De~Blas, G.~Durieux, C.~Grojean, J.~Gu, and A.~Paul, ``{On the future of
  Higgs, electroweak and diboson measurements at lepton colliders},''
  \href{http://dx.doi.org/10.1007/JHEP12(2019)117}{{\em JHEP} {\bfseries 12}
  (2019) 117}, \href{http://arxiv.org/abs/1907.04311}{{\ttfamily
  arXiv:1907.04311 [hep-ph]}}.

\bibitem{Biekoetter:2018ypq}
A.~Biekoetter, T.~Corbett, and T.~Plehn, ``{The Gauge-Higgs Legacy of the LHC
  Run II},'' \href{http://dx.doi.org/10.21468/SciPostPhys.6.6.064}{{\em SciPost
  Phys.} {\bfseries 6} no.~6, (2019) 064},
  \href{http://arxiv.org/abs/1812.07587}{{\ttfamily arXiv:1812.07587
  [hep-ph]}}.

\bibitem{DiVita:2017eyz}
S.~Di~Vita, C.~Grojean, G.~Panico, M.~Riembau, and T.~Vantalon, ``{A global
  view on the Higgs self-coupling},''
  \href{http://dx.doi.org/10.1007/JHEP09(2017)069}{{\em JHEP} {\bfseries 09}
  (2017) 069}, \href{http://arxiv.org/abs/1704.01953}{{\ttfamily
  arXiv:1704.01953 [hep-ph]}}.

\bibitem{deBlas:2017xtg}
J.~de~Blas, J.~C. Criado, M.~Perez-Victoria, and J.~Santiago, ``{Effective
  description of general extensions of the Standard Model: the complete
  tree-level dictionary},''
  \href{http://dx.doi.org/10.1007/JHEP03(2018)109}{{\em JHEP} {\bfseries 03}
  (2018) 109}, \href{http://arxiv.org/abs/1711.10391}{{\ttfamily
  arXiv:1711.10391 [hep-ph]}}.

\bibitem{Perez:1995dc}
M.~A. Perez, J.~J. Toscano, and J.~Wudka, ``{Two photon processes and effective
  Lagrangians with an extended scalar sector},''
  \href{http://dx.doi.org/10.1103/PhysRevD.52.494}{{\em Phys. Rev. D}
  {\bfseries 52} (1995) 494--504},
  \href{http://arxiv.org/abs/hep-ph/9506457}{{\ttfamily arXiv:hep-ph/9506457}}.

\bibitem{Englert:2014uua}
C.~Englert, A.~Freitas, M.~M. Muhlleitner, T.~Plehn, M.~Rauch, M.~Spira, and
  K.~Walz, ``{Precision Measurements of Higgs Couplings: Implications for New
  Physics Scales},''
  \href{http://dx.doi.org/10.1088/0954-3899/41/11/113001}{{\em J. Phys.}
  {\bfseries G41} (2014) 113001},
\href{http://arxiv.org/abs/1403.7191}{{\ttfamily arXiv:1403.7191 [hep-ph]}}.

\bibitem{Brehmer:2015rna}
J.~Brehmer, A.~Freitas, D.~Lopez-Val, and T.~Plehn, ``{Pushing Higgs Effective
  Theory to its Limits},''
  \href{http://dx.doi.org/10.1103/PhysRevD.93.075014}{{\em Phys. Rev. D}
  {\bfseries 93} no.~7, (2016) 075014},
  \href{http://arxiv.org/abs/1510.03443}{{\ttfamily arXiv:1510.03443
  [hep-ph]}}.

\bibitem{Gorbahn:2015gxa}
M.~Gorbahn, J.~M. No, and V.~Sanz, ``{Benchmarks for Higgs Effective Theory:
  Extended Higgs Sectors},''
  \href{http://dx.doi.org/10.1007/JHEP10(2015)036}{{\em JHEP} {\bfseries 10}
  (2015) 036}, \href{http://arxiv.org/abs/1502.07352}{{\ttfamily
  arXiv:1502.07352 [hep-ph]}}.

\bibitem{Belusca-Maito:2016dqe}
H.~B\'elusca-Ma\"\i{}to, A.~Falkowski, D.~Fontes, J.~C. Rom\~ao, and J.~P.
  Silva, ``{Higgs EFT for 2HDM and beyond},''
  \href{http://dx.doi.org/10.1140/epjc/s10052-017-4745-5}{{\em Eur. Phys. J. C}
  {\bfseries 77} no.~3, (2017) 176},
  \href{http://arxiv.org/abs/1611.01112}{{\ttfamily arXiv:1611.01112
  [hep-ph]}}.

\bibitem{Dawson:2017vgm}
S.~Dawson and C.~W. Murphy, ``{Standard Model EFT and Extended Scalar
  Sectors},'' \href{http://dx.doi.org/10.1103/PhysRevD.96.015041}{{\em Phys.
  Rev. D} {\bfseries 96} no.~1, (2017) 015041},
  \href{http://arxiv.org/abs/1704.07851}{{\ttfamily arXiv:1704.07851
  [hep-ph]}}.

\bibitem{Dawson:2020oco}
S.~Dawson, S.~Homiller, and S.~D. Lane, ``{Putting standard model EFT fits to
  work},'' \href{http://dx.doi.org/10.1103/PhysRevD.102.055012}{{\em Phys. Rev.
  D} {\bfseries 102} no.~5, (2020) 055012},
  \href{http://arxiv.org/abs/2007.01296}{{\ttfamily arXiv:2007.01296
  [hep-ph]}}.

\bibitem{Cullen:2020zof}
J.~M. Cullen and B.~D. Pecjak, ``{Higgs decay to fermion pairs at NLO in
  SMEFT},'' \href{http://dx.doi.org/10.1007/JHEP11(2020)079}{{\em JHEP}
  {\bfseries 11} (2020) 079}, \href{http://arxiv.org/abs/2007.15238}{{\ttfamily
  arXiv:2007.15238 [hep-ph]}}.

\bibitem{Cullen:2019nnr}
J.~M. Cullen, B.~D. Pecjak, and D.~J. Scott, ``{NLO corrections to $h\to b\bar
  b$ decay in SMEFT},''
\href{http://arxiv.org/abs/1904.06358}{{\ttfamily arXiv:1904.06358 [hep-ph]}}.

\bibitem{Gauld:2016kuu}
R.~Gauld, B.~D. Pecjak, and D.~J. Scott, ``{QCD radiative corrections for $h\to
  b\bar b$ in the Standard Model Dimension-6 EFT},''
  \href{http://dx.doi.org/10.1103/PhysRevD.94.074045}{{\em Phys. Rev.}
  {\bfseries D94} no.~7, (2016) 074045},
\href{http://arxiv.org/abs/1607.06354}{{\ttfamily arXiv:1607.06354 [hep-ph]}}.

\bibitem{Hartmann:2015aia}
C.~Hartmann and M.~Trott, ``{Higgs Decay to Two Photons at One Loop in the
  Standard Model Effective Field Theory},''
  \href{http://dx.doi.org/10.1103/PhysRevLett.115.191801}{{\em Phys. Rev.
  Lett.} {\bfseries 115} no.~19, (2015) 191801},
\href{http://arxiv.org/abs/1507.03568}{{\ttfamily arXiv:1507.03568 [hep-ph]}}.

\bibitem{Hartmann:2015oia}
C.~Hartmann and M.~Trott, ``{On one-loop corrections in the standard model
  effective field theory; the $\Gamma(h \rightarrow \gamma \, \gamma)$ case},''
  \href{http://dx.doi.org/10.1007/JHEP07(2015)151}{{\em JHEP} {\bfseries 07}
  (2015) 151},
\href{http://arxiv.org/abs/1505.02646}{{\ttfamily arXiv:1505.02646 [hep-ph]}}.

\bibitem{Dawson:2018liq}
S.~Dawson and P.~P. Giardino, ``{Electroweak corrections to Higgs boson decays
  to $\gamma\gamma$ and $W^+W^-$ in standard model EFT},''
  \href{http://dx.doi.org/10.1103/PhysRevD.98.095005}{{\em Phys. Rev.}
  {\bfseries D98} no.~9, (2018) 095005},
\href{http://arxiv.org/abs/1807.11504}{{\ttfamily arXiv:1807.11504 [hep-ph]}}.

\bibitem{Dedes:2018seb}
A.~Dedes, M.~Paraskevas, J.~Rosiek, K.~Suxho, and L.~Trifyllis, ``{The decay
  $h\to \gamma\gamma$ in the Standard-Model Effective Field Theory},''
  \href{http://dx.doi.org/10.1007/JHEP08(2018)103}{{\em JHEP} {\bfseries 08}
  (2018) 103},
\href{http://arxiv.org/abs/1805.00302}{{\ttfamily arXiv:1805.00302 [hep-ph]}}.

\bibitem{Dawson:2018pyl}
S.~Dawson and P.~P. Giardino, ``{Higgs decays to $ZZ$ and $Z\gamma$ in the
  standard model effective field theory: An NLO analysis},''
  \href{http://dx.doi.org/10.1103/PhysRevD.97.093003}{{\em Phys. Rev.}
  {\bfseries D97} no.~9, (2018) 093003},
\href{http://arxiv.org/abs/1801.01136}{{\ttfamily arXiv:1801.01136 [hep-ph]}}.

\bibitem{Dedes:2019bew}
A.~Dedes, K.~Suxho, and L.~Trifyllis, ``{The decay $h\to Z \gamma$ in the
  Standard-Model Effective Field Theory},''
  \href{http://dx.doi.org/10.1007/JHEP06(2019)115}{{\em JHEP} {\bfseries 06}
  (2019) 115},
\href{http://arxiv.org/abs/1903.12046}{{\ttfamily arXiv:1903.12046 [hep-ph]}}.

\bibitem{Dawson:2019clf}
S.~Dawson and P.~P. Giardino, ``{Electroweak and QCD corrections to $Z$ and $W$
  pole observables in the standard model EFT},''
  \href{http://dx.doi.org/10.1103/PhysRevD.101.013001}{{\em Phys. Rev. D}
  {\bfseries 101} no.~1, (2020) 013001},
  \href{http://arxiv.org/abs/1909.02000}{{\ttfamily arXiv:1909.02000
  [hep-ph]}}.

\bibitem{Hartmann:2016pil}
C.~Hartmann, W.~Shepherd, and M.~Trott, ``{The $Z$ decay width in the SMEFT:
  $y_t$ and $\lambda$ corrections at one loop},''
  \href{http://dx.doi.org/10.1007/JHEP03(2017)060}{{\em JHEP} {\bfseries 03}
  (2017) 060},
\href{http://arxiv.org/abs/1611.09879}{{\ttfamily arXiv:1611.09879 [hep-ph]}}.

\bibitem{Boughezal:2019xpp}
R.~Boughezal, C.-Y. Chen, F.~Petriello, and D.~Wiegand, ``{Top quark decay at
  next-to-leading order in the Standard Model Effective Field Theory},''
\href{http://arxiv.org/abs/1907.00997}{{\ttfamily arXiv:1907.00997 [hep-ph]}}.

\bibitem{Dawson:2018dxp}
S.~Dawson, P.~P. Giardino, and A.~Ismail, ``{Standard model EFT and the
  Drell-Yan process at high energy},''
  \href{http://dx.doi.org/10.1103/PhysRevD.99.035044}{{\em Phys. Rev. D}
  {\bfseries 99} no.~3, (2019) 035044},
  \href{http://arxiv.org/abs/1811.12260}{{\ttfamily arXiv:1811.12260
  [hep-ph]}}.

\bibitem{Dawson:2021xea}
S.~Dawson and P.~P. Giardino, ``{New physics through Drell-Yan standard model
  EFT measurements at NLO},''
  \href{http://dx.doi.org/10.1103/PhysRevD.104.073004}{{\em Phys. Rev. D}
  {\bfseries 104} no.~7, (2021) 073004}.

\bibitem{Henning:2016lyp}
B.~Henning, X.~Lu, and H.~Murayama, ``{One-loop Matching and Running with
  Covariant Derivative Expansion},''
  \href{http://dx.doi.org/10.1007/JHEP01(2018)123}{{\em JHEP} {\bfseries 01}
  (2018) 123}, \href{http://arxiv.org/abs/1604.01019}{{\ttfamily
  arXiv:1604.01019 [hep-ph]}}.

\bibitem{DasBakshi:2018vni}
S.~Das~Bakshi, J.~Chakrabortty, and S.~K. Patra, ``{CoDEx: Wilson coefficient
  calculator connecting SMEFT to UV theory},''
  \href{http://dx.doi.org/10.1140/epjc/s10052-018-6444-2}{{\em Eur. Phys. J. C}
  {\bfseries 79} no.~1, (2019) 21},
  \href{http://arxiv.org/abs/1808.04403}{{\ttfamily arXiv:1808.04403
  [hep-ph]}}.

\bibitem{Cohen:2022tir}
T.~Cohen, X.~Lu, and Z.~Zhang, ``{Snowmass White Paper: Effective Field Theory
  Matching and Applications},'' in {\em {2022 Snowmass Summer Study}}.
\newblock 3, 2022.
\newblock \href{http://arxiv.org/abs/2203.07336}{{\ttfamily arXiv:2203.07336
  [hep-ph]}}.

\bibitem{Carmona:2021xtq}
A.~Carmona, A.~Lazopoulos, P.~Olgoso, and J.~Santiago, ``{Matchmakereft:
  automated tree-level and one-loop matching},''
  \href{http://arxiv.org/abs/2112.10787}{{\ttfamily arXiv:2112.10787
  [hep-ph]}}.

\bibitem{Criado:2017khh}
J.~C. Criado, ``{MatchingTools: a Python library for symbolic effective field
  theory calculations},''
  \href{http://dx.doi.org/10.1016/j.cpc.2018.02.016}{{\em Comput. Phys.
  Commun.} {\bfseries 227} (2018) 42--50},
  \href{http://arxiv.org/abs/1710.06445}{{\ttfamily arXiv:1710.06445
  [hep-ph]}}.

\bibitem{Jiang:2018pbd}
M.~Jiang, N.~Craig, Y.-Y. Li, and D.~Sutherland, ``{Complete one-loop matching
  for a singlet scalar in the Standard Model EFT},''
  \href{http://dx.doi.org/10.1007/JHEP02(2019)031}{{\em JHEP} {\bfseries 02}
  (2019) 031}, \href{http://arxiv.org/abs/1811.08878}{{\ttfamily
  arXiv:1811.08878 [hep-ph]}}. [Erratum: JHEP 01, 135 (2021)].

\bibitem{Haisch:2020ahr}
U.~Haisch, M.~Ruhdorfer, E.~Salvioni, E.~Venturini, and A.~Weiler, ``{Singlet
  night in Feynman-ville: one-loop matching of a real scalar},''
  \href{http://dx.doi.org/10.1007/JHEP04(2020)164}{{\em JHEP} {\bfseries 04}
  (2020) 164}, \href{http://arxiv.org/abs/2003.05936}{{\ttfamily
  arXiv:2003.05936 [hep-ph]}}. [Erratum: JHEP 07, 066 (2020)].

\bibitem{Dawson:2021jcl}
S.~Dawson, P.~P. Giardino, and S.~Homiller, ``{Uncovering the High Scale Higgs
  Singlet Model},'' \href{http://dx.doi.org/10.1103/PhysRevD.103.075016}{{\em
  Phys. Rev. D} {\bfseries 103} no.~7, (2021) 075016},
  \href{http://arxiv.org/abs/2102.02823}{{\ttfamily arXiv:2102.02823
  [hep-ph]}}.

\bibitem{Cohen:2020xca}
T.~Cohen, N.~Craig, X.~Lu, and D.~Sutherland, ``{Is SMEFT Enough?},''
  \href{http://dx.doi.org/10.1007/JHEP03(2021)237}{{\em JHEP} {\bfseries 03}
  (2021) 237}, \href{http://arxiv.org/abs/2008.08597}{{\ttfamily
  arXiv:2008.08597 [hep-ph]}}.

\bibitem{Anisha:2021hgc}
Anisha, S.~Das~Bakshi, S.~Banerjee, A.~Biek\"otter, J.~Chakrabortty,
  S.~Kumar~Patra, and M.~Spannowsky, ``{Effective limits on single scalar
  extensions in the light of recent LHC data},''
  \href{http://arxiv.org/abs/2111.05876}{{\ttfamily arXiv:2111.05876
  [hep-ph]}}.

\bibitem{Hays:2018zze}
C.~Hays, A.~Martin, V.~Sanz, and J.~Setford, ``{On the impact of
  dimension-eight SMEFT operators on Higgs measurements},''
  \href{http://dx.doi.org/10.1007/JHEP02(2019)123}{{\em JHEP} {\bfseries 02}
  (2019) 123}, \href{http://arxiv.org/abs/1808.00442}{{\ttfamily
  arXiv:1808.00442 [hep-ph]}}.

\bibitem{Corbett:2021eux}
T.~Corbett, A.~Helset, A.~Martin, and M.~Trott, ``{EWPD in the SMEFT to
  dimension eight},'' \href{http://dx.doi.org/10.1007/JHEP06(2021)076}{{\em
  JHEP} {\bfseries 06} (2021) 076},
  \href{http://arxiv.org/abs/2102.02819}{{\ttfamily arXiv:2102.02819
  [hep-ph]}}.

\bibitem{Dawson:2021xei}
S.~Dawson, S.~Homiller, and M.~Sullivan, ``{Impact of dimension-eight SMEFT
  contributions: A case study},''
  \href{http://dx.doi.org/10.1103/PhysRevD.104.115013}{{\em Phys. Rev. D}
  {\bfseries 104} no.~11, (2021) 115013},
  \href{http://arxiv.org/abs/2110.06929}{{\ttfamily arXiv:2110.06929
  [hep-ph]}}.

\bibitem{Lee:1973iz}
T.~D. Lee, ``{A Theory of Spontaneous T Violation},''
\href{http://dx.doi.org/10.1103/PhysRevD.8.1226}{{\em Phys. Rev.} {\bfseries
  D8} (1973) 1226--1239}.

\bibitem{Gunion:1989we}
J.~F. Gunion, H.~E. Haber, G.~L. Kane, and S.~Dawson, ``{The Higgs Hunter's
  Guide},''
{\em Front. Phys.} {\bfseries 80} (2000) 1--404.

\bibitem{Branco:2011iw}
G.~C. Branco, P.~M. Ferreira, L.~Lavoura, M.~N. Rebelo, M.~Sher, and J.~P.
  Silva, ``{Theory and phenomenology of two-Higgs-doublet models},''
  \href{http://dx.doi.org/10.1016/j.physrep.2012.02.002}{{\em Phys. Rept.}
  {\bfseries 516} (2012) 1--102},
\href{http://arxiv.org/abs/1106.0034}{{\ttfamily arXiv:1106.0034 [hep-ph]}}.

\bibitem{Belusca-Maito:2017iob}
H.~B\'elusca-Ma\"\i{}to, A.~Falkowski, D.~Fontes, J.~C. Rom\~ao, and J.~P.
  Silva, ``{CP violation in 2HDM and EFT: the $ZZZ$ vertex},''
  \href{http://dx.doi.org/10.1007/JHEP04(2018)002}{{\em JHEP} {\bfseries 04}
  (2018) 002}, \href{http://arxiv.org/abs/1710.05563}{{\ttfamily
  arXiv:1710.05563 [hep-ph]}}.

\bibitem{Donoghue:1978cj}
J.~F. Donoghue and L.~F. Li, ``{Properties of Charged Higgs Bosons},''
  \href{http://dx.doi.org/10.1103/PhysRevD.19.945}{{\em Phys. Rev. D}
  {\bfseries 19} (1979) 945}.

\bibitem{Georgi:1978ri}
H.~Georgi and D.~V. Nanopoulos, ``{Suppression of Flavor Changing Effects From
  Neutral Spinless Meson Exchange in Gauge Theories},''
  \href{http://dx.doi.org/10.1016/0370-2693(79)90433-7}{{\em Phys. Lett. B}
  {\bfseries 82} (1979) 95--96}.

\bibitem{Botella:1994cs}
F.~J. Botella and J.~P. Silva, ``{Jarlskog - like invariants for theories with
  scalars and fermions},''
  \href{http://dx.doi.org/10.1103/PhysRevD.51.3870}{{\em Phys. Rev.} {\bfseries
  D51} (1995) 3870--3875},
\href{http://arxiv.org/abs/hep-ph/9411288}{{\ttfamily arXiv:hep-ph/9411288
  [hep-ph]}}.

\bibitem{Branco:1999fs}
G.~C. Branco, L.~Lavoura, and J.~P. Silva, ``{CP Violation},''
{\em Int. Ser. Monogr. Phys.} {\bfseries 103} (1999) 1--536.

\bibitem{Fontes:2021znm}
D.~Fontes, M.~L\"oschner, J.~C. Rom\~ao, and J.~P. Silva, ``{Leaks of CP
  violation in the real two-Higgs-doublet model},''
  \href{http://dx.doi.org/10.1140/epjc/s10052-021-09332-0}{{\em Eur. Phys. J.
  C} {\bfseries 81} no.~6, (2021) 541},
  \href{http://arxiv.org/abs/2103.05002}{{\ttfamily arXiv:2103.05002
  [hep-ph]}}.

\bibitem{Belusca-Maito:2016cay}
H.~B\'elusca-Ma\"\i{}to, {\em {Search for new physics at the LHC using Higgs
  Effective Field Theory}}.
\newblock PhD thesis, Orsay, 3, 2016.

\bibitem{Henning:2014wua}
B.~Henning, X.~Lu, and H.~Murayama, ``{How to use the Standard Model effective
  field theory},'' \href{http://dx.doi.org/10.1007/JHEP01(2016)023}{{\em JHEP}
  {\bfseries 01} (2016) 023}, \href{http://arxiv.org/abs/1412.1837}{{\ttfamily
  arXiv:1412.1837 [hep-ph]}}.

\bibitem{Egana-Ugrinovic:2015vgy}
D.~Egana-Ugrinovic and S.~Thomas, ``{Effective Theory of Higgs Sector Vacuum
  States},'' \href{http://arxiv.org/abs/1512.00144}{{\ttfamily arXiv:1512.00144
  [hep-ph]}}.

\bibitem{Buchmuller:1985jz}
W.~Buchmuller and D.~Wyler, ``{Effective Lagrangian Analysis of New
  Interactions and Flavor Conservation},''
  \href{http://dx.doi.org/10.1016/0550-3213(86)90262-2}{{\em Nucl. Phys. B}
  {\bfseries 268} (1986) 621--653}.

\bibitem{Grzadkowski:2010es}
B.~Grzadkowski, M.~Iskrzynski, M.~Misiak, and J.~Rosiek, ``{Dimension-Six Terms
  in the Standard Model Lagrangian},''
  \href{http://dx.doi.org/10.1007/JHEP10(2010)085}{{\em JHEP} {\bfseries 10}
  (2010) 085}, \href{http://arxiv.org/abs/1008.4884}{{\ttfamily arXiv:1008.4884
  [hep-ph]}}.

\bibitem{Murphy:2020rsh}
C.~W. Murphy, ``{Dimension-8 operators in the Standard Model Eective Field
  Theory},'' \href{http://dx.doi.org/10.1007/JHEP10(2020)174}{{\em JHEP}
  {\bfseries 10} (2020) 174}, \href{http://arxiv.org/abs/2005.00059}{{\ttfamily
  arXiv:2005.00059 [hep-ph]}}.

\bibitem{Li:2020xlh}
H.-L. Li, Z.~Ren, M.-L. Xiao, J.-H. Yu, and Y.-H. Zheng, ``{Complete set of
  dimension-nine operators in the standard model effective field theory},''
  \href{http://dx.doi.org/10.1103/PhysRevD.104.015025}{{\em Phys. Rev. D}
  {\bfseries 104} no.~1, (2021) 015025},
  \href{http://arxiv.org/abs/2007.07899}{{\ttfamily arXiv:2007.07899
  [hep-ph]}}.

\bibitem{Criado:2018sdb}
J.~C. Criado and M.~P\'erez-Victoria, ``{Field redefinitions in effective
  theories at higher orders},''
  \href{http://dx.doi.org/10.1007/JHEP03(2019)038}{{\em JHEP} {\bfseries 03}
  (2019) 038}, \href{http://arxiv.org/abs/1811.09413}{{\ttfamily
  arXiv:1811.09413 [hep-ph]}}.

\bibitem{Gunion:2002zf}
J.~F. Gunion and H.~E. Haber, ``{The CP conserving two Higgs doublet model: The
  Approach to the decoupling limit},''
  \href{http://dx.doi.org/10.1103/PhysRevD.67.075019}{{\em Phys. Rev. D}
  {\bfseries 67} (2003) 075019},
  \href{http://arxiv.org/abs/hep-ph/0207010}{{\ttfamily arXiv:hep-ph/0207010}}.

\bibitem{Ferreira:2014naa}
P.~M. Ferreira, J.~F. Gunion, H.~E. Haber, and R.~Santos, ``{Probing wrong-sign
  Yukawa couplings at the LHC and a future linear collider},''
  \href{http://dx.doi.org/10.1103/PhysRevD.89.115003}{{\em Phys. Rev.}
  {\bfseries D89} no.~11, (2014) 115003},
\href{http://arxiv.org/abs/1403.4736}{{\ttfamily arXiv:1403.4736 [hep-ph]}}.

\bibitem{ParticleDataGroup:2020ssz}
{\bfseries Particle Data Group} Collaboration, P.~A. Zyla {\em et~al.},
  ``{Review of Particle Physics},''
  \href{http://dx.doi.org/10.1093/ptep/ptaa104}{{\em PTEP} {\bfseries 2020}
  no.~8, (2020) 083C01}.

\bibitem{https://doi.org/10.23731/cyrm-2017-002}
{CERN}, ``Cern yellow reports: Monographs, vol 2 (2017): Handbook of lhc higgs
  cross sections: 4. deciphering the nature of the higgs sector,'' 2017.
\newblock \url{https://e-publishing.cern.ch/index.php/CYRM/issue/view/32}.

\bibitem{ATLAS:2016neq}
{\bfseries ATLAS, CMS} Collaboration, G.~Aad {\em et~al.}, ``{Measurements of
  the Higgs boson production and decay rates and constraints on its couplings
  from a combined ATLAS and CMS analysis of the LHC pp collision data at $
  \sqrt{s}=7 $ and 8 TeV},''
  \href{http://dx.doi.org/10.1007/JHEP08(2016)045}{{\em JHEP} {\bfseries 08}
  (2016) 045}, \href{http://arxiv.org/abs/1606.02266}{{\ttfamily
  arXiv:1606.02266 [hep-ex]}}.

\bibitem{ATLAS:2020qdt}
{\bfseries ATLAS} Collaboration, ``{A combination of measurements of Higgs
  boson production and decay using up to $139$ fb$^{-1}$ of proton--proton
  collision data at $\sqrt{s}=$ 13 TeV collected with the ATLAS experiment},''.

\bibitem{CMS:2020gsy}
{\bfseries CMS} Collaboration, ``{Combined Higgs boson production and decay
  measurements with up to 137 fb$^{-1}$ of proton-proton collision data at
  $\sqrt s$ = 13 TeV},''.

\bibitem{Degrassi:2016wml}
G.~Degrassi, P.~P. Giardino, F.~Maltoni, and D.~Pagani, ``{Probing the Higgs
  self coupling via single Higgs production at the LHC},''
  \href{http://dx.doi.org/10.1007/JHEP12(2016)080}{{\em JHEP} {\bfseries 12}
  (2016) 080}, \href{http://arxiv.org/abs/1607.04251}{{\ttfamily
  arXiv:1607.04251 [hep-ph]}}.

\bibitem{Degrassi:2021uik}
G.~Degrassi, B.~Di~Micco, P.~P. Giardino, and E.~Rossi, ``{Higgs boson
  self-coupling constraints from single Higgs, double Higgs and Electroweak
  measurements},'' \href{http://dx.doi.org/10.1016/j.physletb.2021.136307}{{\em
  Phys. Lett. B} {\bfseries 817} (2021) 136307},
  \href{http://arxiv.org/abs/2102.07651}{{\ttfamily arXiv:2102.07651
  [hep-ph]}}.

\bibitem{Atkinson:2021eox}
O.~Atkinson, M.~Black, A.~Lenz, A.~Rusov, and J.~Wynne, ``{Cornering the Two
  Higgs Doublet Model Type II},''
  \href{http://arxiv.org/abs/2107.05650}{{\ttfamily arXiv:2107.05650
  [hep-ph]}}.

\bibitem{Contino:2016jqw}
R.~Contino, A.~Falkowski, F.~Goertz, C.~Grojean, and F.~Riva, ``{On the
  Validity of the Effective Field Theory Approach to SM Precision Tests},''
  \href{http://dx.doi.org/10.1007/JHEP07(2016)144}{{\em JHEP} {\bfseries 07}
  (2016) 144}, \href{http://arxiv.org/abs/1604.06444}{{\ttfamily
  arXiv:1604.06444 [hep-ph]}}.

\end{thebibliography}\endgroup
\end{document}